\documentclass[showpacs]{revtex4}

\usepackage{graphicx}
\usepackage{wrapft}
\usepackage{amsmath,amssymb}
\usepackage{bm}
\usepackage{color}
\usepackage{textcomp}
\usepackage{multirow}

\begin{document}

\title{Cluster structures in $^{11}$B}

\author{Tadahiro Suhara}

\affiliation{Yukawa Institute for Theoretical Physics, Kyoto University, Kyoto 606-8502, Japan}

\author{Yoshiko Kanada-En'yo}

\affiliation{Department of Physics, Graduate School of Science, Kyoto University, Kyoto 606-8502, Japan}

\begin{abstract}
Structures of excited states in $^{11}$B are investigated with a method of 
$\beta$-$\gamma$ constraint antisymmetrized molecular dynamics in combination with the generator coordinate method.
Various excited states with developed cluster core structures are suggested in positive- and negative-parity states.
For negative-parity states, we suggest a band with a $2\alpha$+$t$ cluster structure.
This band starts from the $3/2^{-}_{3}$ state and can correspond to the experimental band observed recently.
In positive-parity states, two $\alpha$ core cluster structures with surrounding nucleons are found. 
A $K^\pi=1/2^+$ band is suggested to be constructed from a remarkably developed cluster structure with a large prolate deformation.
We discuss features of the cluster structure in association with molecular orbital structures of $^{10}$Be.
\end{abstract}

\pacs{21.60.-n, 02.70.Ns, 21.10.Ky, 27.20.+n}

\maketitle

\section{Introduction}\label{introduction}

Cluster aspect is one of the essential features as well as shell-model aspect. 
In light nuclei, cluster and shell-model features often coexist as discussed, 
for example, for 3$\alpha$ cluster structures in $^{12}$C \cite{Horiuchi_OCM_74,Uegaki_12C_77,Kamimura_12C_77,Descouvemont_12C_87,En'yo_12C_98,Tohsaki_12C_01,Funaki_12C_03,Neff_12C_04,Funaki_12C_05,En'yo_12C_07,Kurokawa_12C_07}.

In the recent experiment of $\alpha$ resonant scattering on $^{7}$Li \cite{Yamaguchi_11B_11},
a new negative-parity band consisting of 
8.56 MeV ($3/2^{-}$), 10.34 MeV ($5/2^{-}$), 11.59 MeV ($7/2^{-}$), and 13.03 MeV ($9/2^{-}$) was suggested.
Since these states have large $\alpha$ decay widths,
this band is considered to be a band constructed from a cluster structure.
$^{11}$B can be an interesting nucleus where cluster and shell-model structures coexist. 
Indeed, it was suggested in previous works that
low-lying states of $^{11}$B have shell-model structure mainly \cite{Navratil_NCSM_03}, 
while, cluster structures develop well in the negative-parity states 
above or near the threshold \cite{Nishioka_11B_79,En'yo_11B_07,Yamada_11B_10}.

Moreover, analogy of cluster features in $^{11}$B with three $\alpha$ cluster structures in $^{12}$C 
is an fascinating problem to be clarified \cite{En'yo_11B_07}.
In the previous works \cite{En'yo_11B_07,Kawabata_11B_07}, the $3/2^{-}_{3}$ state was suggested 
to have a dilute cluster structure with a $2\alpha + t$ configuration and to be an analog state with 
the $0^{+}_{2}$ state in $^{12}$C, which has a dilute 3$\alpha$ structure.
However, in the recent work \cite{Yamada_11B_10}, it is suggested that 
$^{11}$B($3/2^{-}_{3}$) cannot correspond to $^{12}$C($0^{+}_{2}$). 
The relation between $^{11}$B($3/2^{-}_{3}$) and $^{12}$C($0^{+}_{2}$) is controversial and
further studies for this problem are required.

For the positive-parity states of $^{11}$B, there are few theoretical studies 
though cluster states are expected to appear near the threshold energy.
Thus, structures of excited states of $^{11}$B is a challenging problem to study.

In this article, we investigate structures of excited states in $^{11}$B with 
a method of $\beta$-$\gamma$ constraint antisymmetrized molecular dynamics (AMD) 
in combination with the generator coordinate method (GCM).
To clarify the correspondence between the $3/2^{-}_{3}$ state of $^{11}$B and the $0^{+}_{2}$ state of $^{12}$C, 
we compare their GCM amplitudes on the $\beta$-$\gamma$ plane.
We also discuss molecular orbital structures with a $2\alpha$ core and surrounding nucleons for positive-parity states in $^{11}$B 
and their correspondence with $^{10}$Be.

This paper is organized as follows. 
In Sec. \ref{framework}, we explain the framework of the $\beta$-$\gamma$ constraint AMD + GCM briefly.
The calculated results are shown in Sec. \ref{results}.
In Sec. \ref{discussion}, we give discussions about structures of $^{11}$B.
Finally, in Sec. \ref{summary}, we summarize this paper.

\section{Framework}\label{framework}

The frameworks of AMD are described in detail, for example, in Refs.~\cite{En'yo_PTP_95,En'yo_AMD_95,En'yo_sup_01,En'yo_AMD_03}.
In the present work, we adopt a version of AMD, the $\beta$-$\gamma$ constraint AMD \cite{Suhara_AMD_10}, 
in which we perform the variation with the constraint on the quadrupole deformation parameters, $\beta$ and $\gamma$.

In the method of AMD, 
a basis wave function of an $A$-nucleon system $|\Phi \rangle$ 
is described by a Slater determinant of single-particle wave functions $|\varphi_{i} \rangle$ as
\begin{equation}
|\Phi \rangle = \frac{1}{\sqrt{A!}} \det \left\{ |\varphi_{1} \rangle, \cdots ,|\varphi_{A} \rangle \right\}.
\end{equation}
The $i$-th single-particle wave function $|\varphi_{i} \rangle$ consists of 
the spatial part $|\phi_{i} \rangle$, spin part $|\chi_{i} \rangle$, and isospin part $|\tau_{i} \rangle$ as
\begin{equation}
	|\varphi_{i} \rangle = |\phi_{i} \rangle |\chi_{i} \rangle |\tau_{i} \rangle.
\end{equation}
The spatial part $|\phi_{i} \rangle$ is given by a Gaussian wave packet
whose center is located at $\bm{Z}_{i}/\sqrt{\nu}$ as
\begin{equation}
	\langle \bm{r} | \phi_{i} \rangle = \left( \frac{2\nu}{\pi} \right)^{\frac{3}{4}}
		\exp \left[ - \nu \left( \bm{r} - \frac{\bm{Z}_{i}}{\sqrt{\nu}} \right)^{2} 
		+ \frac{1}{2} \bm{Z}_{i}^{2}\right] 
	\label{single_particle_spatial}, 
\end{equation}
where $\nu$ is the width parameter and is taken to be a common value for all the
single-particle Gaussian wave functions in the present work.
The spin orientation is given by the parameter $\bm{\xi}_{i}$, while
the isospin part $|\tau_{i} \rangle$ is fixed to be up (proton) or down (neutron), 
\begin{align}
	|\chi_{i} \rangle &= \xi_{i\uparrow} |\uparrow \ \rangle + \xi_{i\downarrow} |\downarrow \ \rangle,\\
	|\tau_{i} \rangle &= |p \rangle \ \text{or} \ |n \rangle.
\end{align}
In a basis wave function $|\Phi \rangle$, $\{ X \} \equiv \{ \bm{Z} , \bm{\xi} \} = \{ \bm{Z}_{1} , \bm{\xi}_{1} , \bm{Z}_{2} , \bm{\xi}_{2} , 
\cdots , \bm{Z}_{A} , \bm{\xi}_{A} \}$ are complex variational parameters and they 
are determined by the energy optimization.

We perform the variation for the parity projected wave function $|\Phi ^{\pm} \rangle$ defined as
\begin{equation}
	|\Phi ^{\pm} \rangle \equiv \hat{P}^{\pm} |\Phi \rangle,
\end{equation}
with the constraint on the quadrupole deformation parameters, $\beta$ and $\gamma$
to obtain various cluster and shell-model structures as the basis wave functions.
The definition of $\beta$ and $\gamma$ are
\begin{align}
	&\beta \cos \gamma \equiv \frac{\sqrt{5\pi}}{3} 
		\frac{2\langle \hat{z}^{2} \rangle -\langle \hat{x}^{2} \rangle -\langle \hat{y}^{2} \rangle }{R^{2}}, \\
	&\beta \sin \gamma \equiv \sqrt{\frac{5\pi}{3}} 
		\frac{\langle \hat{x}^{2} \rangle -\langle \hat{y}^{2} \rangle }{R^{2}} \label{definition_beta_gamma}, \\
	&R^{2} \equiv \frac{5}{3} \left( \langle \hat{x}^{2} \rangle + \langle \hat{y}^{2} \rangle 
		+ \langle \hat{z}^{2} \rangle \right).
\end{align}
Here, $\langle \hat{O} \rangle$ represents the expectation value of the operator $\hat{O}$ for an intrinsic wave function $| \Phi \rangle$.
After the variation with the constraints, we obtain the optimized wave functions
$|\Phi^{\pm}(\beta_{0}, \gamma_{0}) \rangle$
for each set of parameters, $(\beta, \gamma) = (\beta_{0}, \gamma_{0})$.

In the calculations of energy levels, 
we superpose the parity and total-angular-momentum projected 
AMD wave functions $\hat{P}^{J}_{MK} |\Phi^{\pm}(\beta, \gamma) \rangle$ using GCM. 
Thus, the final wave function for the $J^\pm_n$ state is given by
a linear combination of the basis wave functions as 
\begin{equation}
	|\Phi ^{J\pm}_{n} \rangle = \sum_{K} \sum_{i} f_{n}(\beta_{i}, \gamma_{i}, K) \hat{P}^{J}_{MK} |\Phi^{\pm}(\beta_{i}, \gamma_{i}) \rangle.
	\label{dispersed_GCM}
\end{equation}
The coefficients $f_{n}(\beta_{i}, \gamma_{i}, K)$ are determined using the Hill-Wheeler equation.

For the effective two-body interactions,
we use the Volkov No.~2 interaction \cite{Volkov_No2_65} as the central force
and the spin-orbit term of the G3RS interaction \cite{G3RS_79} as the $LS$ force.
We take the same interaction parameters as those in Refs.~\cite{Suhara_AMD_10,Suhara_14C_10,Suhara_linear-chain_11}, i.e.,
the Majorana exchange parameter $M = 0.6$ ($W = 0.4$), the Bartlett exchange parameter $B = 0.125$,
and the Heisenberg exchange parameter $H = 0.125$ in the central force, and 
$u_{1} = -1600$ MeV and $u_{2} = 1600$ MeV in the $LS$ force.
These parameters are the same as those adopted in the studies for $^{9}$Be \cite{Okabe_9Be_79}, 
and $^{10}$Be \cite{Itagaki_10Be_00}, except for a small modification in the strength of the spin-orbit force
to fit the $0^{+}_{1}$ energy of $^{12}$C \cite{Suhara_AMD_10}.

For the width parameter of single-particle Gaussian wave packets in Eq.~\eqref{single_particle_spatial}, 
we used the value $\nu = 0.235$ fm$^{-2}$, which is also the same as those in the studies for C isotopes 
\cite{Suhara_AMD_10,Suhara_14C_10,Itagaki_Cisotopes_01,Itagaki_14C_04}.

\section{Results}\label{results}

We performed variational calculations with the $\beta$-$\gamma$ constraint 
at 196 mesh points of the triangle lattice on the $\beta$-$\gamma$ plane
and superposed the obtained wave functions. 
In this section, we show the calculated results.

\subsection{Energy surfaces}

\begin{figure}[t]
	\centering
	\begin{tabular}{c}
	\includegraphics[width=8.6cm, bb=10 12 700 364, clip]{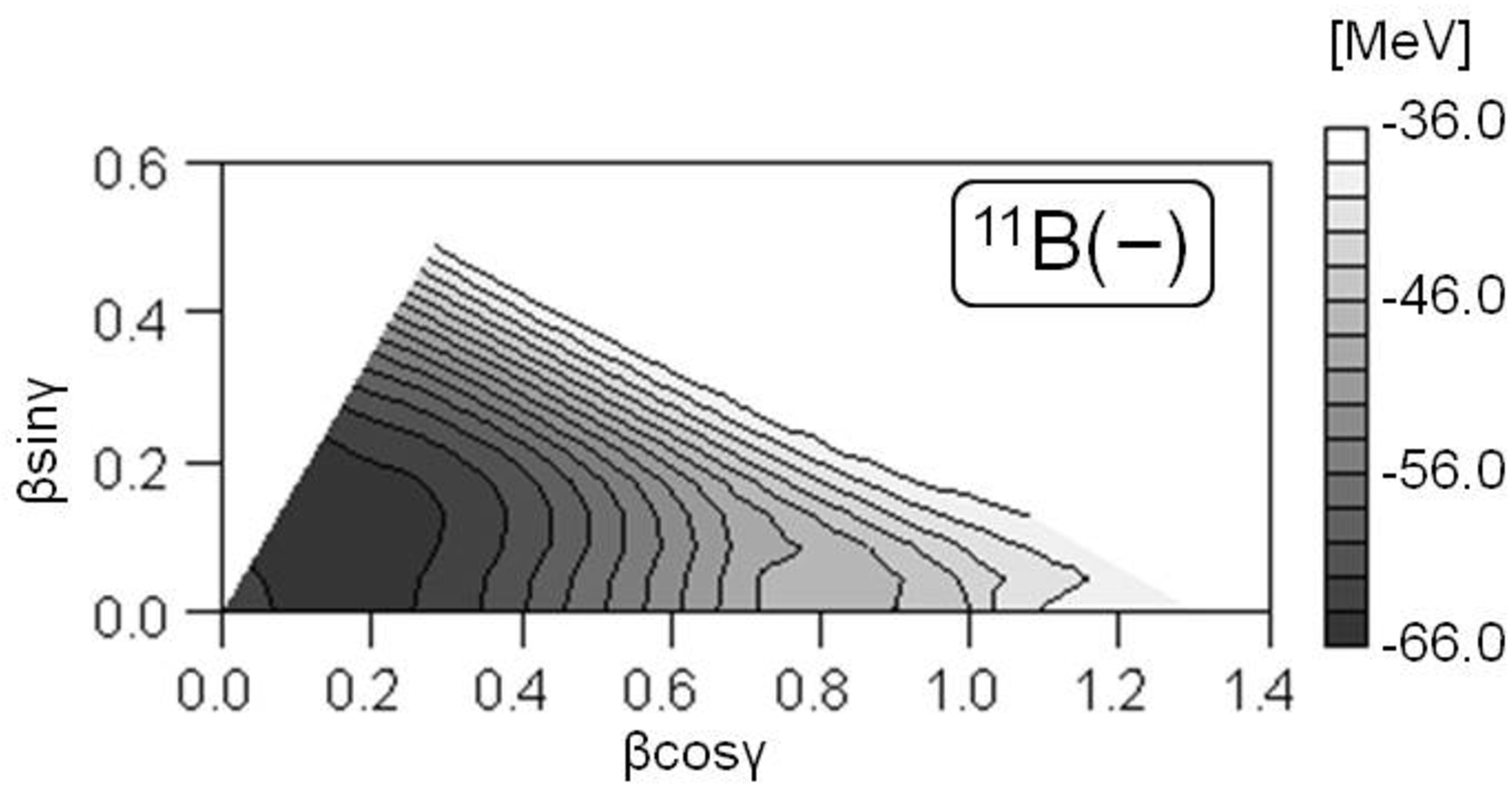} \\
	\includegraphics[width=8.6cm, bb=10 12 700 364, clip]{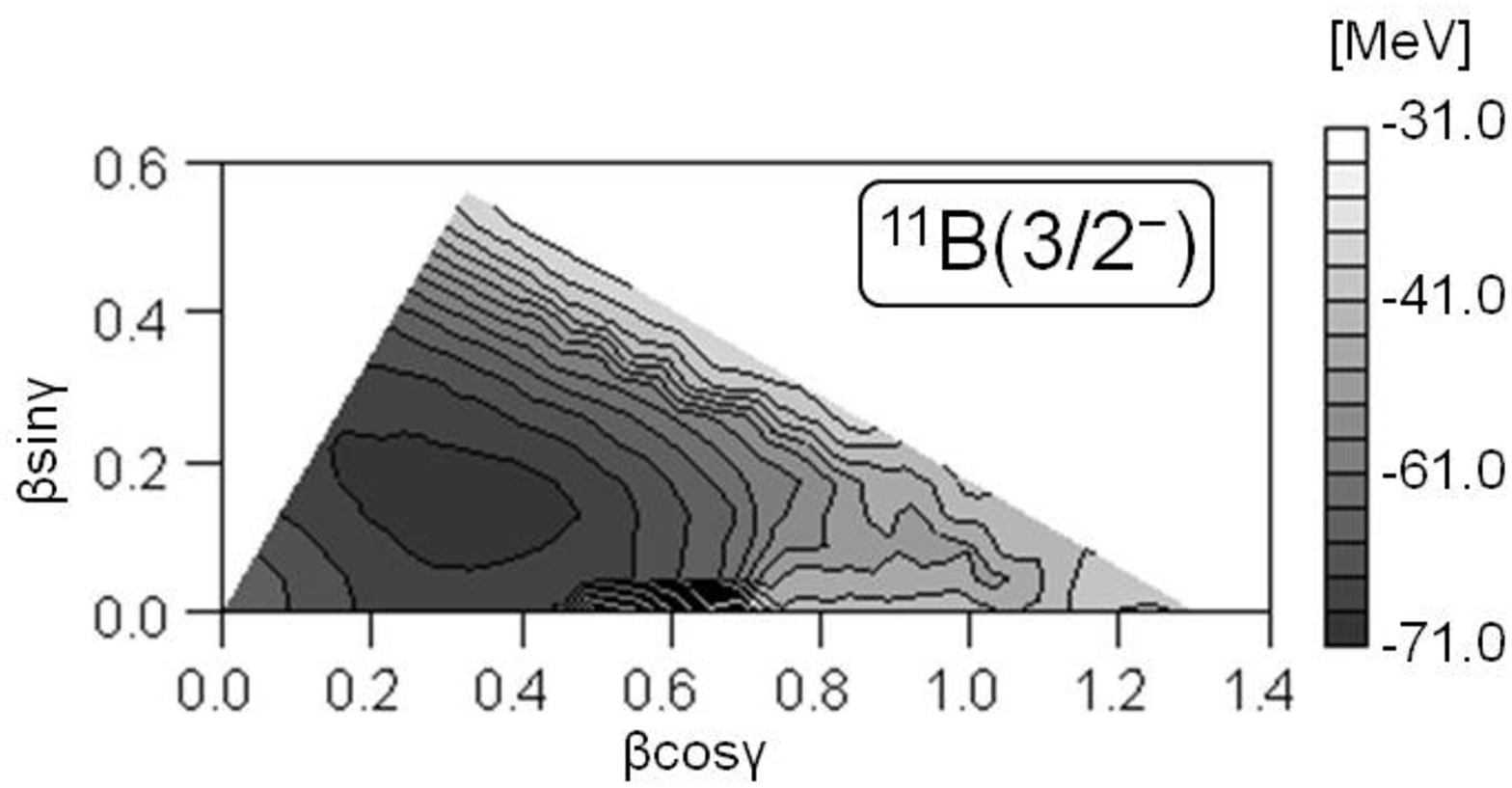}
	\end{tabular}
	\caption{Energy surfaces of $^{11}$B on the $\beta$-$\gamma$ plane.
	The top panel shows the energy for the negative-parity states and 
	the bottom panel shows that for the $3/2^{-}$ states after the total-angular-momentum projection.}
	\label{11B_energy_surface_-}
\end{figure}

\begin{figure}[t]
	\begin{tabular}{c}
	\includegraphics[width=8.6cm, bb=10 12 700 364, clip]{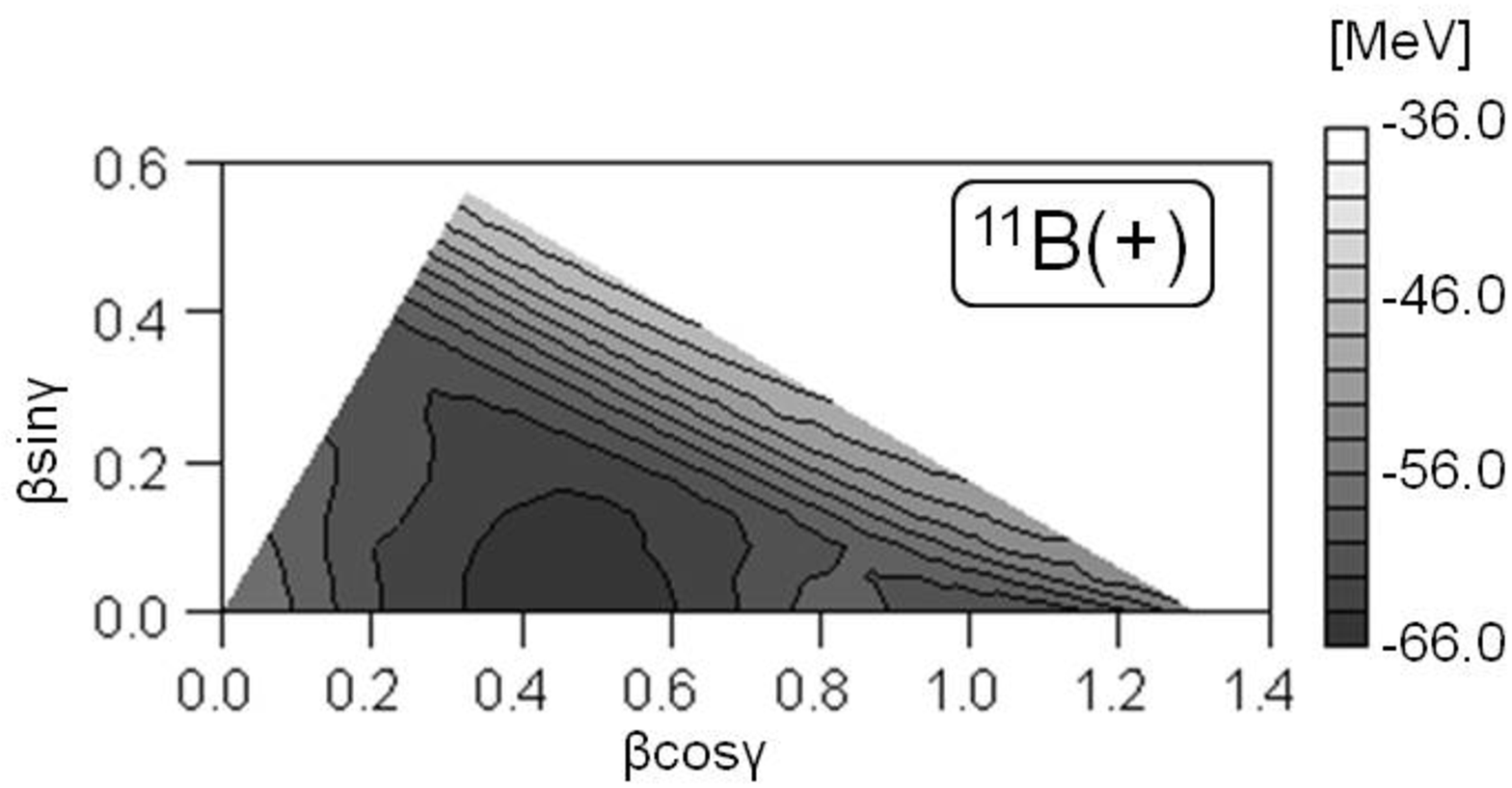} \\
	\includegraphics[width=8.6cm, bb=10 12 700 364, clip]{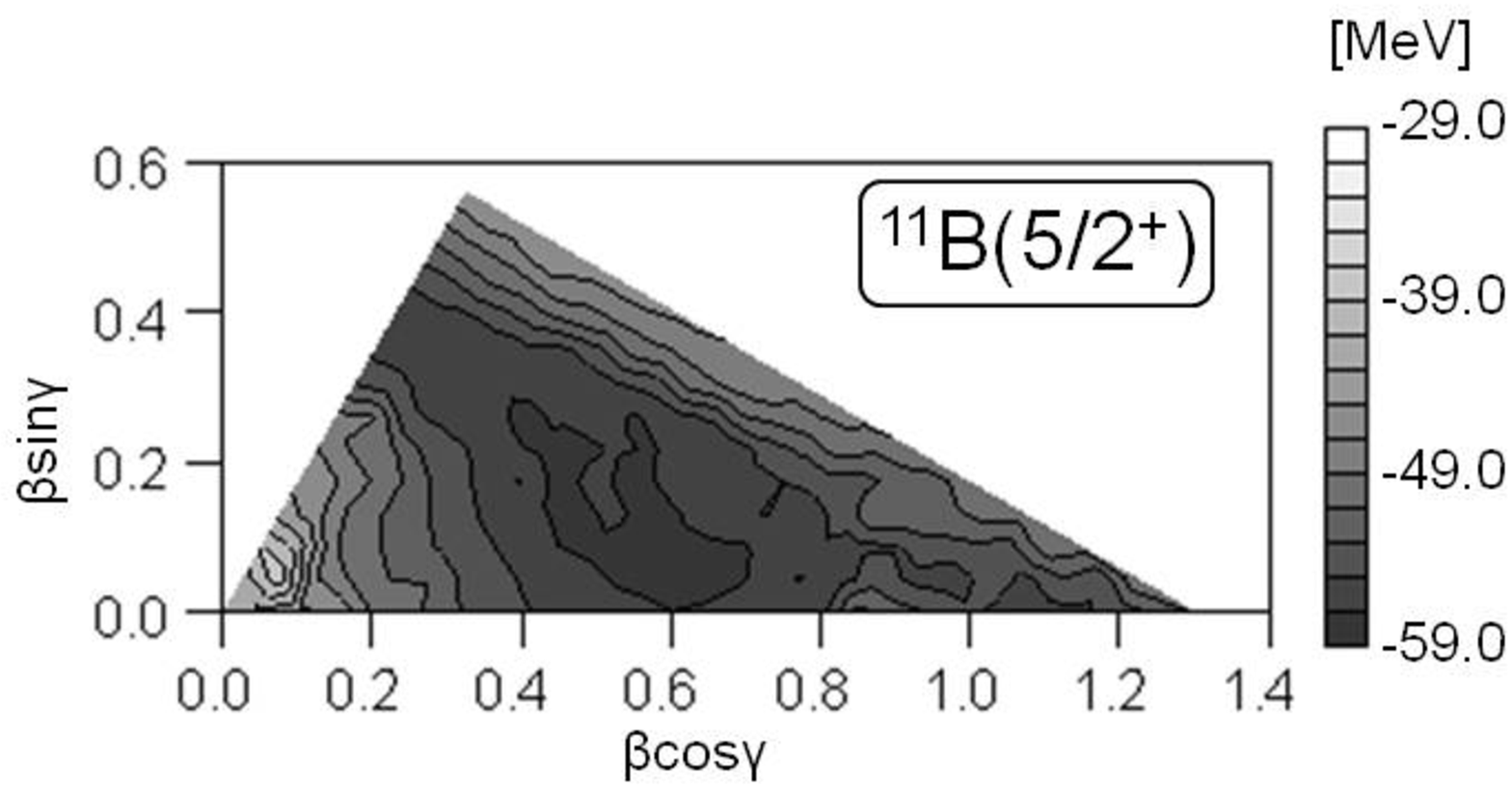} 
	\end{tabular}
	\caption{Energy surfaces of $^{11}$B on the $\beta$-$\gamma$ plane.
	The top panel shows the energy for the positive-parity states and 
	the bottom panel shows that for the $5/2^{+}$ states after the total-angular-momentum projection.}
	\label{11B_energy_surface_+}
\end{figure}

Energy surfaces as functions of $\beta$ and $\gamma$ are obtained.
The calculated energy surfaces for negative-parity states and those for positive-parity states are shown 
in Figs.~\ref{11B_energy_surface_-} and \ref{11B_energy_surface_+}, respectively.

In Fig.~\ref{11B_energy_surface_-}, the top panel  shows the energy surface for the negative-parity states
and the bottom panel shows that for the $3/2^{-}$ states after the total-angular-momentum projection.
We call the former the negative-parity energy surface and the latter the $3/2^{-}$ energy surface.
The minimum point of negative-parity energy surface is at $(\beta \cos \gamma, \beta \sin \gamma) = (0.13, 0.13)$.
After the total-angular-momentum projection onto $3/2^{-}$ eigenstates, 
the minimum point shifts to $(\beta \cos \gamma, \beta \sin \gamma) = (0.33, 0.13)$. 
This indicates that the deformation of the energy minimum state becomes large after the total-angular-momentum projection.
In the large prolate region, a valley is found around $(\beta \cos \gamma, \beta \sin \gamma) = (0.9, 0.1)$.
Interestingly, the feature of the  $3/2^{-}$ energy surface for $^{11}$B is similar to 
that of the $0^{+}$ energy surface for $^{12}$C (See Fig.~2 in Ref.~\cite{Suhara_AMD_10}).

The positive-parity energy surface and the $5/2^{+}$ energy surface 
are displayed in the top and bottom panels of Fig.~\ref{11B_energy_surface_+}, respectively.
The minimum point of the positive-parity energy surface is at
$(\beta \cos \gamma, \beta \sin \gamma) = (0.45, 0.00)$.
After the total-angular-momentum projection onto $5/2^{+}$ eigenstates, 
the minimum point shifts to $(\beta \cos \gamma, \beta \sin \gamma) = (0.60, 0.09)$. 
Thus the deformation of the energy minimum state changes 
from the prolate deformation before the total-angular-momentum projection 
to the large $\beta$ and triaxial region after the projection.
In a largely deformed region, a local minimum exists at $(\beta \cos \gamma, \beta \sin \gamma) = (1.00, 0.00)$ 
in the positive-parity energy surface and it is at $(\beta \cos \gamma, \beta \sin \gamma) = (1.10, 0.00)$
in the $5/2^{+}$ energy surface.
As we show later, a rotational band with the large prolate deformation 
is constructed by wave functions in this region after the GCM calculation.

\subsection{Structures on the $\beta$-$\gamma$ plane}

In this section, we explain the intrinsic structures of negative- and positive-parity states obtained 
with the $\beta$-$\gamma$ constraint AMD.

We analyze the spatial configurations of the Gaussian centers $\{\bm{Z}_1,\bm{Z}_2,\cdots,\bm{Z}_A\}$ 
and the distributions of proton density $\rho_{p}$, neutron density $\rho_{n}$ and 
the neutron-proton density difference $\rho_{n} - \rho_{p}$ of each intrinsic wave function $|\Phi(\beta, \gamma) \rangle$
obtained for given constraint values, $\beta$ and $\gamma$.
The neutron-proton density difference $\rho_{n} - \rho_{p}$ show excess neutron behaviors.
We show density distributions $\Tilde{\rho}$ which are integrated densities along the $y$-axis as 
\begin{align}
	\Tilde{\rho} (x, z) &\equiv \int dy \rho (\bm{r}), \\ 
	\rho (\bm{r}) &\equiv \langle \Phi (\beta, \gamma) | \sum_{i} \delta (\bm{r} - \hat{\bm{r}}_{i}) |\Phi (\beta, \gamma) \rangle.
	\label{density_distribution}
\end{align}

\begin{figure}[tb]
	\centering
	{\tabcolsep=0.7mm
	\begin{tabular}{ccccc}
	\vspace{0.3cm}
	($\beta \cos \gamma$, $\beta \sin \gamma$) &
	\hspace{-0.2cm} $\Tilde{\rho}_{p}$ & \hspace{-0.2cm} $\Tilde{\rho}_{n}$ & \hspace{-0.15cm} $\Tilde{\rho}_{n} - \Tilde{\rho}_{p}$ & [1/fm$^{2}$] \\
	\vspace{-2.2cm}
	\begin{minipage}{.17\linewidth}\vspace{-1.7cm}(a) ($0.33$, $0.13$)\end{minipage} &
	\includegraphics[width=1.8cm, bb=223 311 394 481, clip]{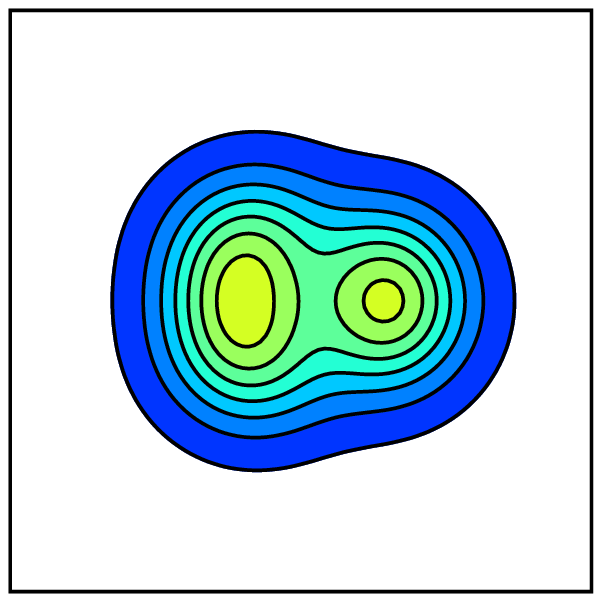} &
	\includegraphics[width=1.8cm, bb=223 311 394 481, clip]{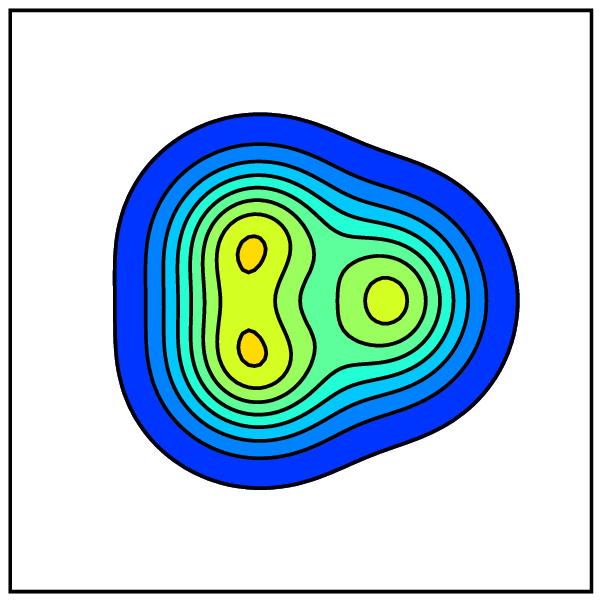} &
	\includegraphics[width=1.8cm, bb=223 311 394 481, clip]{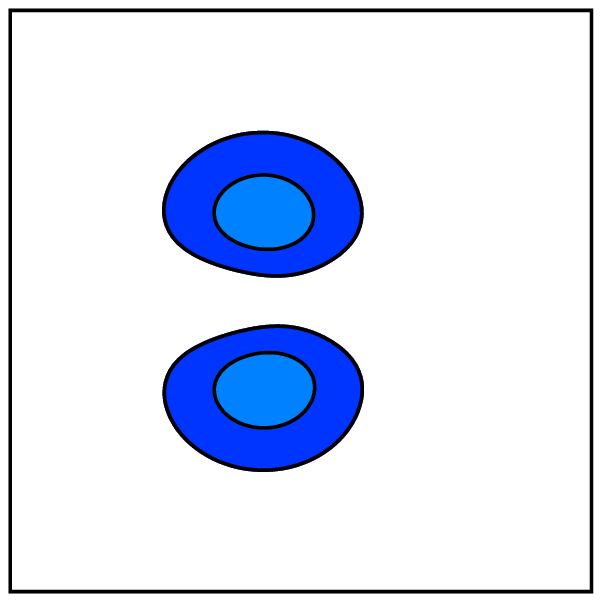} & 
	\begin{minipage}{.10\linewidth}\vspace{-4.4cm}[1/fm$^{2}$]\end{minipage} \\
	\begin{minipage}{.17\linewidth}\vspace{-1.7cm}(b) ($0.90$, $0.09$)\end{minipage} &
	\includegraphics[width=1.8cm, bb=223 311 394 481, clip]{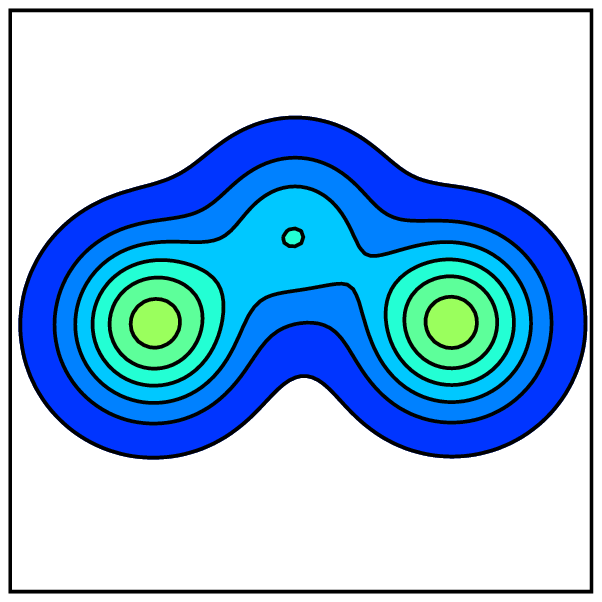} &
	\includegraphics[width=1.8cm, bb=223 311 394 481, clip]{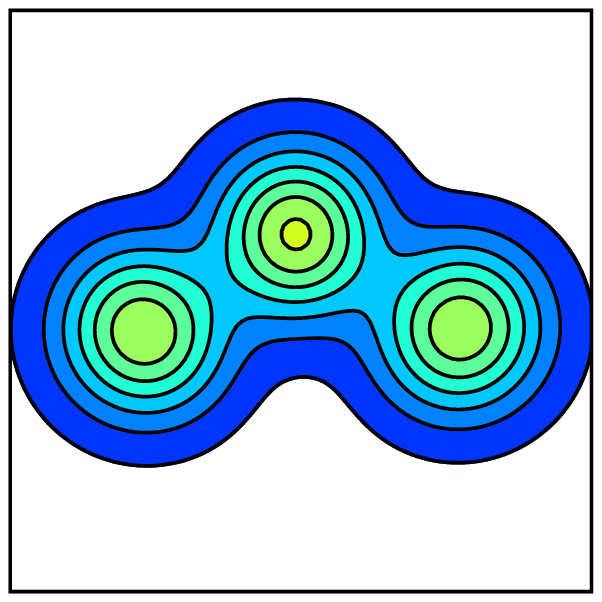} &
	\includegraphics[width=1.8cm, bb=223 311 394 481, clip]{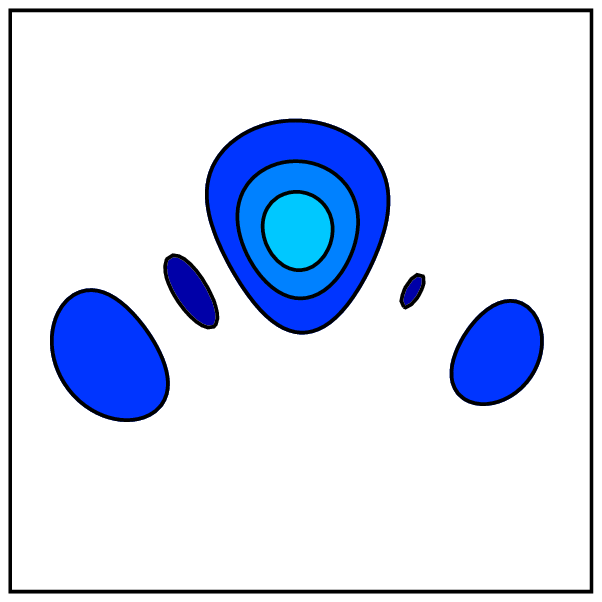} &
	\includegraphics[width=1.5cm, bb=379 301 450 490, clip]{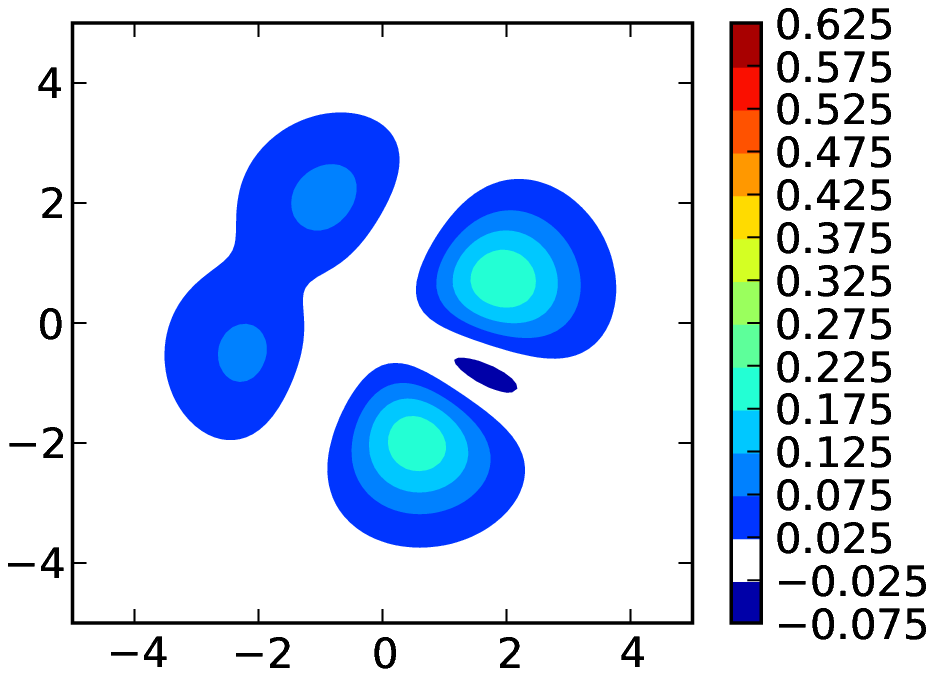} \\
	\begin{minipage}{.17\linewidth}\vspace{-1.7cm}(c) ($0.33$, $0.48$)\end{minipage} &
	\includegraphics[width=1.8cm, bb=223 311 394 481, clip]{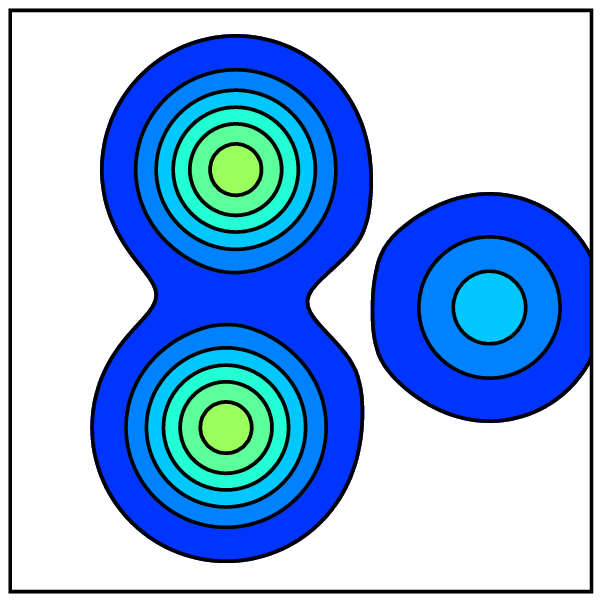} &
	\includegraphics[width=1.8cm, bb=223 311 394 481, clip]{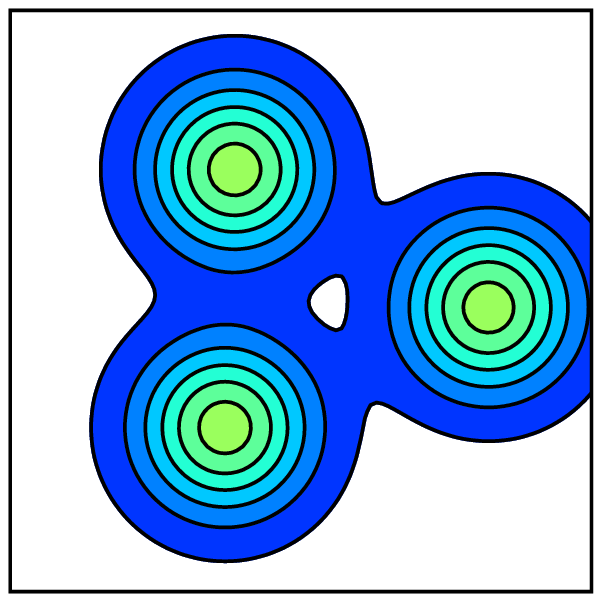} &
	\includegraphics[width=1.8cm, bb=223 311 394 481, clip]{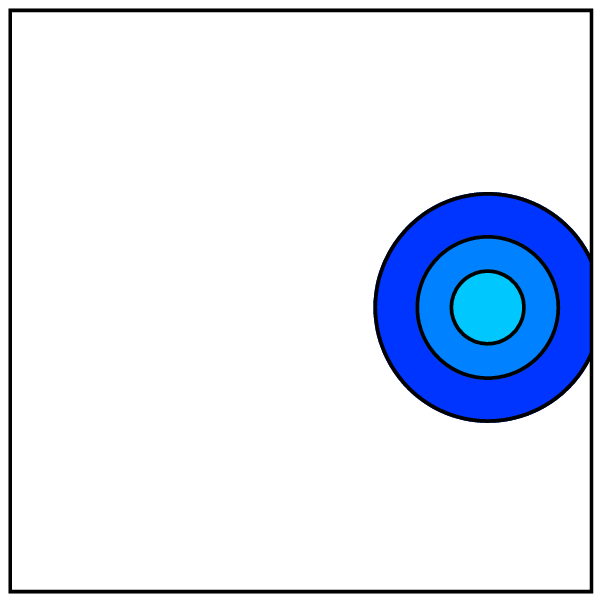} \\
	\begin{minipage}{.17\linewidth}\vspace{-1.7cm}(d) ($0.75$, $0.17$)\end{minipage} &
	\includegraphics[width=1.8cm, bb=223 311 394 481, clip]{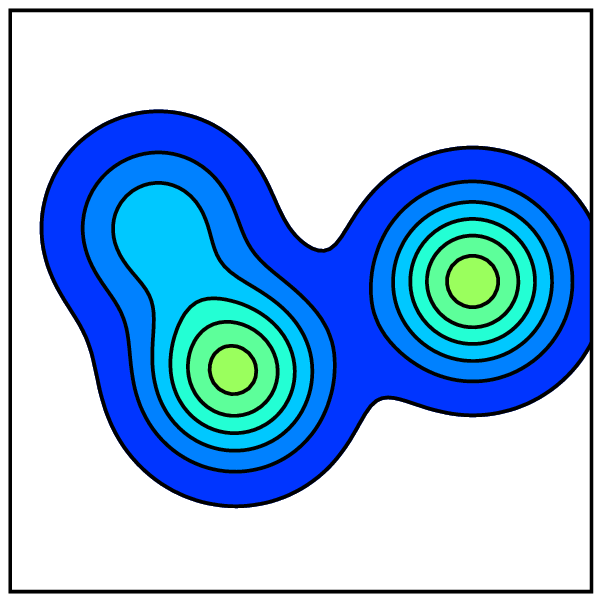} &
	\includegraphics[width=1.8cm, bb=223 311 394 481, clip]{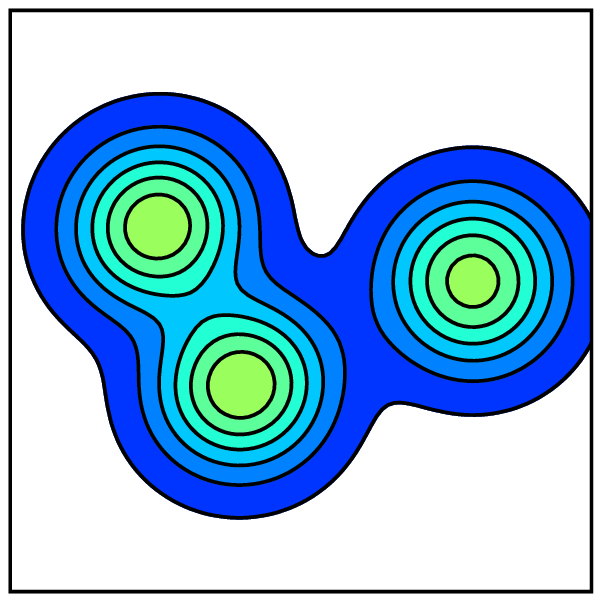} &
	\includegraphics[width=1.8cm, bb=223 311 394 481, clip]{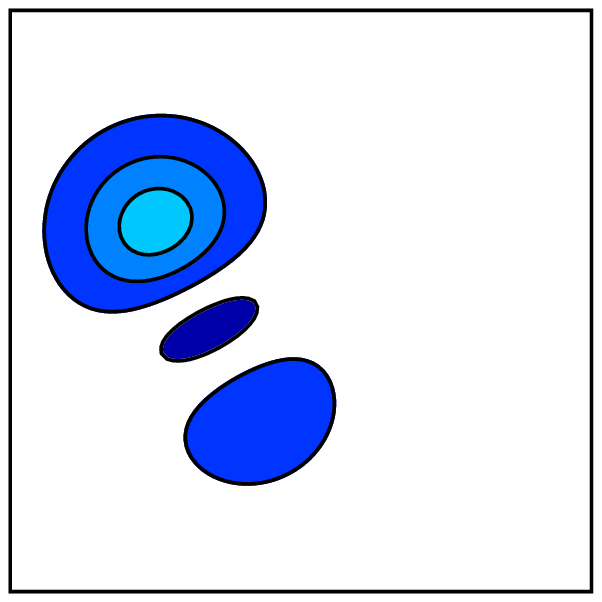} \\
	\end{tabular}}
	\caption{Density distributions of the intrinsic wave functions for the negative-parity states of $^{11}$B.
	The proton density $\Tilde{\rho}_{p}$, neutron density $\Tilde{\rho}_{n}$, 
	and difference between the neutron and proton densities 
	$\Tilde{\rho}_{n} - \Tilde{\rho}_{p}$ are illustrated in the left, middle, and right columns, respectively.
	The density distributions of the intrinsic wave functions at 
	(a) $(\beta \cos \gamma,\beta \sin \gamma)=(0.33,0.13)$,
	(b) $(\beta \cos \gamma,\beta \sin \gamma)=(0.90,0.09)$,
	(c) $(\beta \cos \gamma,\beta \sin \gamma)=(0.33,0.48)$, and 
	(d) $(\beta \cos \gamma,\beta \sin \gamma)=(0.75,0.17)$ on the $\beta$-$\gamma$ plane are shown. 
	The size of the box is 10 $\times$ 10 fm$^{2}$.}
	\label{density_11B_-}
\end{figure}

The density distributions of the intrinsic wave functions for negative-parity states 
are illustrated in Fig.~\ref{density_11B_-}.
Figure~\ref{density_11B_-}(a) is the density distribution for the energy minimum state 
in the $3/2^{-}$ energy surface $(\beta \cos \gamma, \beta \sin \gamma) = (0.33, 0.13)$.
In this wave function, the neutron density has 
a three peak structure showing some components of a $2\alpha + t$ cluster structure,
though spatial development of the clustering is weak as indicated by the fact that 
centers of single-particle Gaussian wave packets gather around the origin.
The expectation value of squared intrinsic spin of neutrons is 0.42, which is an intermediate value 
between 0 for the $2\alpha + t$ cluster limit and 4/3 for the $p_{3/2}$-shell closed configuration limit. 
This result indicates a mixture of the $p_{3/2}$-shell closed configuration and a $2\alpha + t$ cluster structure. 
That is to say, this state is considered to be the intermediate between the shell-model structure and the cluster structure.

In the large deformation region, two $\alpha$ and $t$ clusters develop well. 
Various configurations of clusters appear, depending on the deformation parameters, $\beta$ and $\gamma$. 
Figures~\ref{density_11B_-}(b), (c), and (d) are typical density distributions for prolate, oblate, and
triaxial deformed states, respectively. 
It is found that the linear-chainlike, equilateral-triangular, and obtuse-angle-triangular configuration arise
in the prolate state (b), oblate state (c), and triaxial state (d), respectively.

\begin{figure}[tb]
	\centering
	{\tabcolsep=0.7mm
	\begin{tabular}{ccccc}
	\vspace{0.3cm}
	($\beta \cos \gamma$, $\beta \sin \gamma$) &
	\hspace{-0.2cm} $\Tilde{\rho}_{p}$ & \hspace{-0.2cm} $\Tilde{\rho}_{n}$ & \hspace{-0.15cm} $\Tilde{\rho}_{n} - \Tilde{\rho}_{p}$ \\
	\vspace{-2.2cm}
	\begin{minipage}{.17\linewidth}\vspace{-1.7cm}(a) $(0.60, 0.09)$\end{minipage} &
	\includegraphics[width=1.8cm, bb=223 311 394 481, clip]{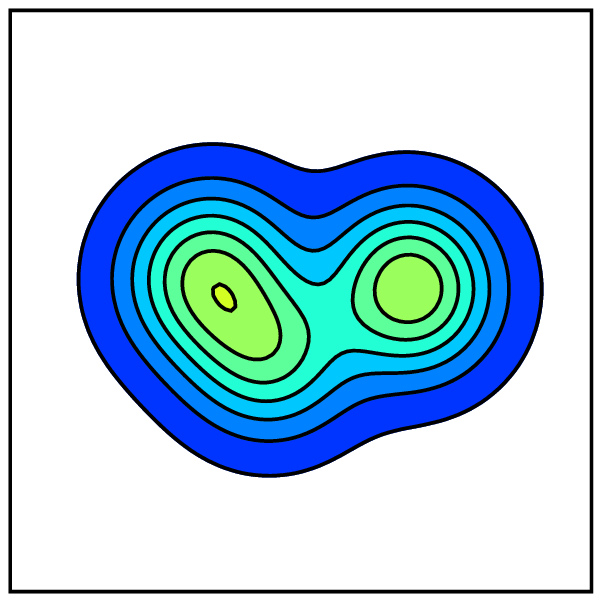} &
	\includegraphics[width=1.8cm, bb=223 311 394 481, clip]{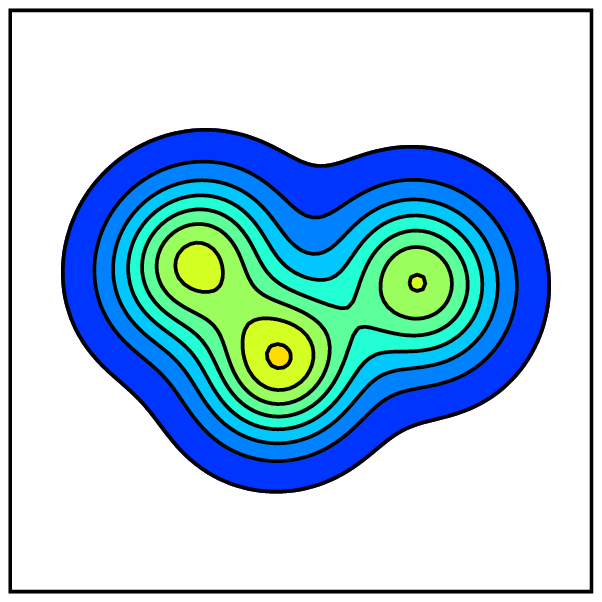} &
	\includegraphics[width=1.8cm, bb=223 311 394 481, clip]{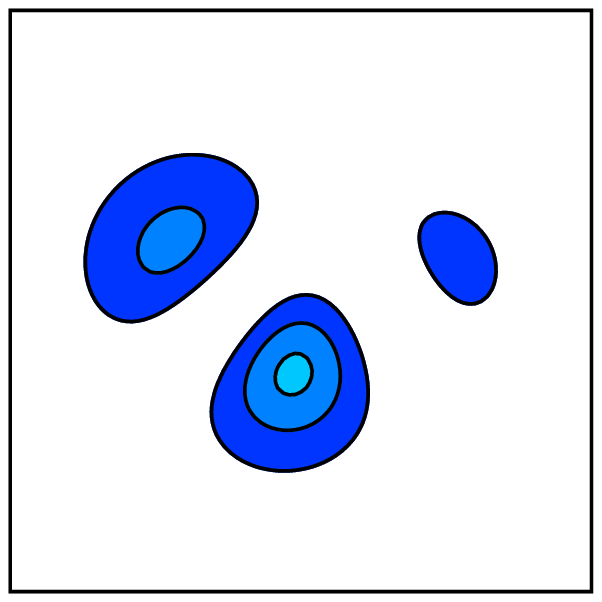} & 
	\begin{minipage}{.10\linewidth}\vspace{-4.4cm}[1/fm$^{2}$]\end{minipage} \\
	\begin{minipage}{.17\linewidth}\vspace{-1.7cm}(b) $(1.10, 0.00)$\end{minipage} &
	\includegraphics[width=1.8cm, bb=223 311 394 481, clip]{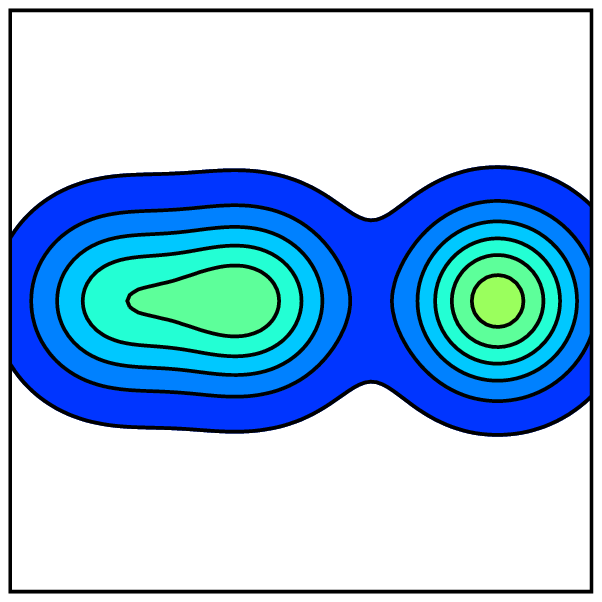} &
	\includegraphics[width=1.8cm, bb=223 311 394 481, clip]{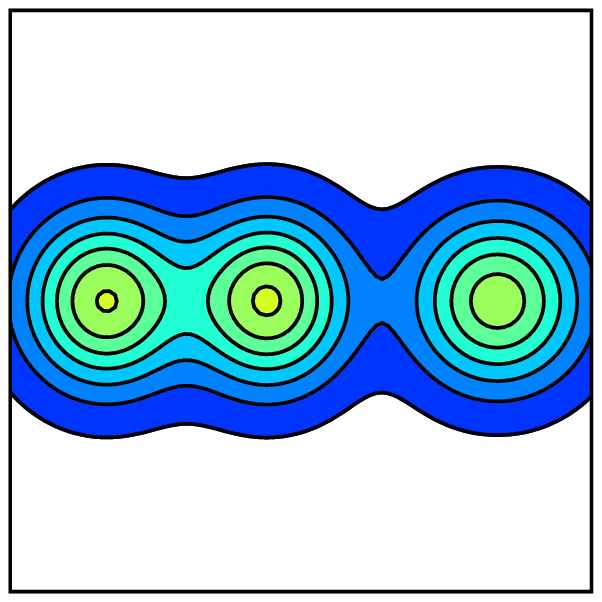} &
	\includegraphics[width=1.8cm, bb=223 311 394 481, clip]{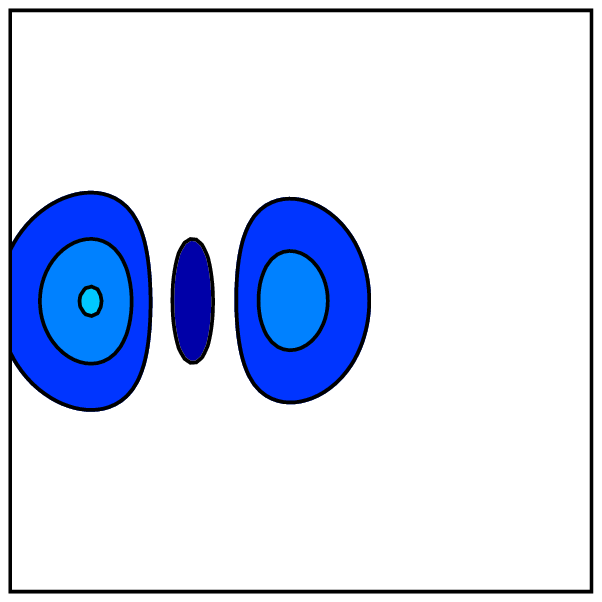} &
	\includegraphics[width=1.5cm, bb=379 301 450 490, clip]{legend_density_+.eps} \\
	\begin{minipage}{.17\linewidth}\vspace{-1.7cm}(c) $(0.28, 0.48)$\end{minipage} &
	\includegraphics[width=1.8cm, bb=223 311 394 481, clip]{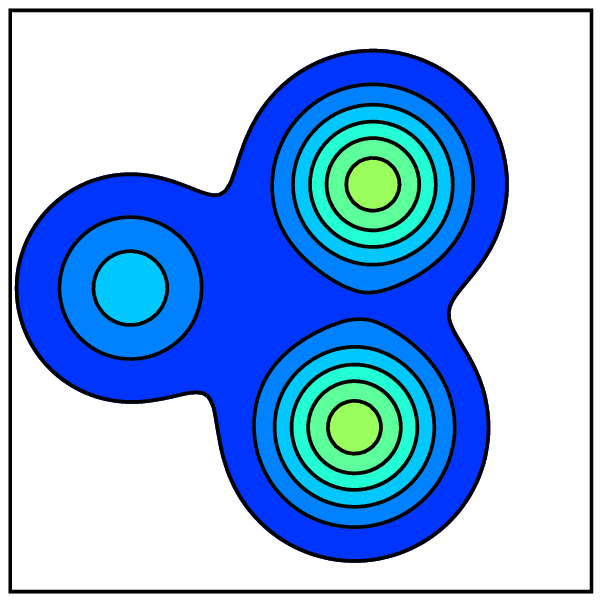} &
	\includegraphics[width=1.8cm, bb=223 311 394 481, clip]{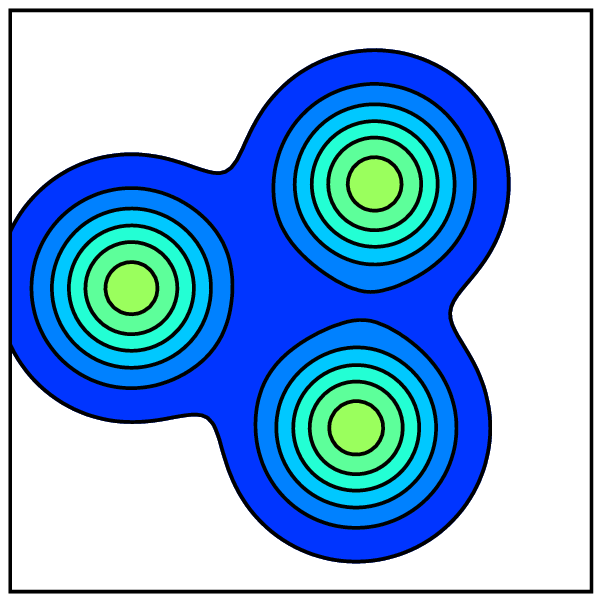} &
	\includegraphics[width=1.8cm, bb=223 311 394 481, clip]{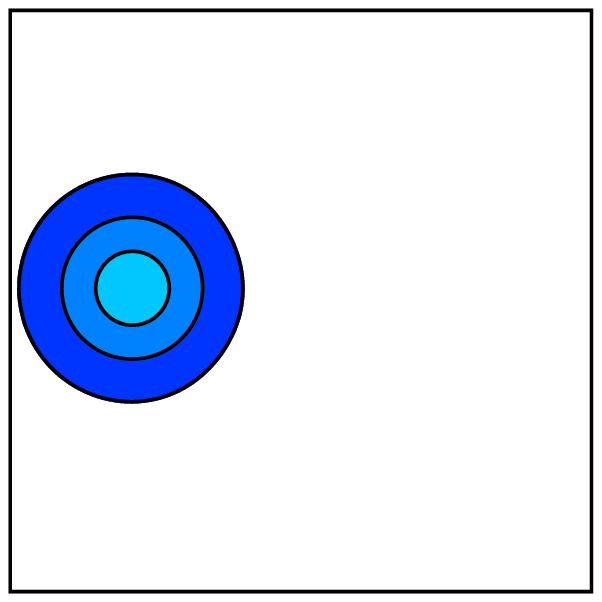} \\
	\begin{minipage}{.17\linewidth}\vspace{-1.7cm}(d) $(0.83, 0.22)$\end{minipage} &
	\includegraphics[width=1.8cm, bb=223 311 394 481, clip]{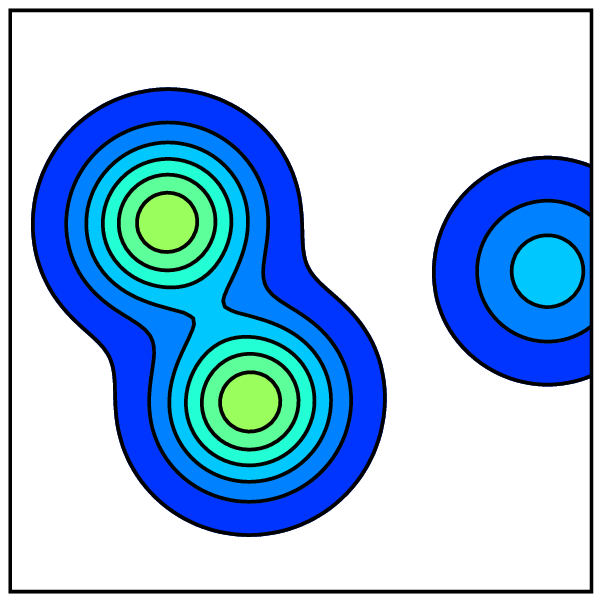} &
	\includegraphics[width=1.8cm, bb=223 311 394 481, clip]{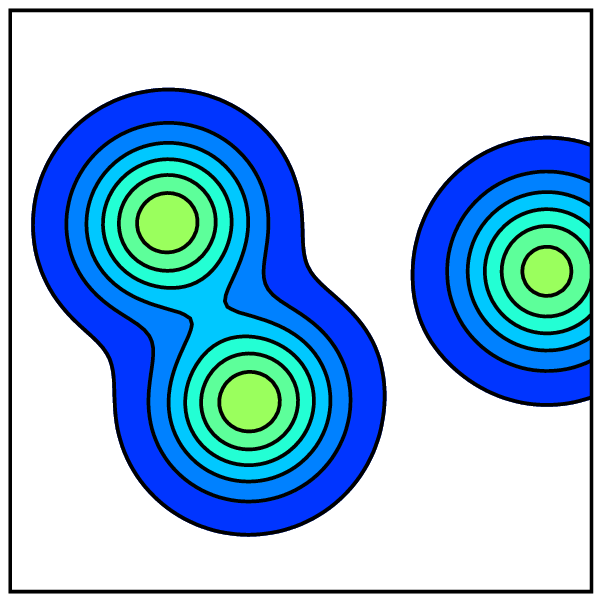} &
	\includegraphics[width=1.8cm, bb=223 311 394 481, clip]{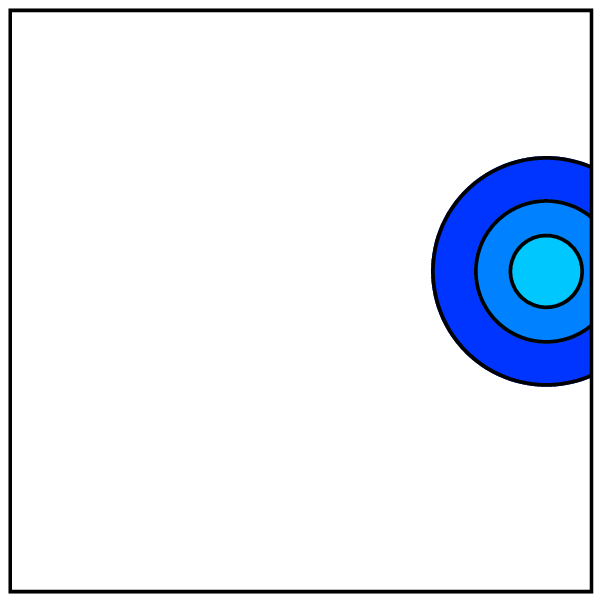} \\
	\end{tabular}}
	\caption{Density distributions of the intrinsic wave functions for the positive-parity states of $^{11}$B.
	The proton density $\Tilde{\rho}_{p}$, neutron density $\Tilde{\rho}_{n}$, 
	and difference between the neutron and proton densities 
	$\Tilde{\rho}_{n} - \Tilde{\rho}_{p}$ are illustrated in the left, middle, and right columns, respectively.
	The density distributions of the intrinsic wave functions at 
	(a) $(\beta \cos \gamma,\beta \sin \gamma)=(0.60,0.09)$,
	(b) $(\beta \cos \gamma,\beta \sin \gamma)=(1.10,0.00)$,
	(c) $(\beta \cos \gamma,\beta \sin \gamma)=(0.28,0.48)$, and 
	(d) $(\beta \cos \gamma,\beta \sin \gamma)=(0.83,0.22)$ on the $\beta$-$\gamma$ plane are shown. 
	The size of the box is 10 $\times$ 10 fm$^{2}$.}
	\label{density_11B_+}
\end{figure}

Various cluster structures are also found in positive-parity states as well as in negative-parity states.
The density distributions of the intrinsic wave functions for positive-parity states 
are illustrated in Fig.~\ref{density_11B_+}.
Figure~\ref{density_11B_+}(a) shows the density distributions 
at $(\beta \cos \gamma, \beta \sin \gamma) = (0.60, 0.09)$, which is 
the energy minimum point of the $5/2^{+}$ energy surface.
In this state, a $2\alpha$ core is formed.
This is seen by the expectation value of the squared proton spin $\langle \hat{S}_{p}^{2} \rangle$ is 0.77, 
which is consistent with the ideal value $\langle \hat{S}_{p}^{2} \rangle = 3/4$ for a 2$\alpha$+$p$+2$n$ configuration
where spins of four protons in two $\alpha$ clusters couple to $S=0$ and a valence proton gives spin $1/2$.

Figure~\ref{density_11B_+}(b) shows the density distribution of the local minimum state
with $(\beta \cos \gamma, \beta \sin \gamma)=(1.10, 0.00)$ in the $5/2^{+}$ energy surface.
In this state, to be clear from the expectation value of the squared proton spin $\langle \hat{S}_{p}^{2} \rangle = 0.80$,
two $\alpha$ clusters are also formed.
Valence nucleons, one proton and two neutrons, 
distribute around one of the two $\alpha$ cluster resulting in the $^{7}$Li correlation.
After the GCM calculation, a $K^{\pi}=1/2^{+}$ rotational band is constructed from this state as discussed later.

Figures~\ref{density_11B_+}(c) and (d) are the density distributions for 
typical structures with oblate and triaxial deformations in the large $\beta$ region.
In these states, $2\alpha + t$ cluster structures develop well.
It is found that the isosceles-triangular structure and obtuse-angle-triangular structure 
of two $\alpha$ and one $t$ clusters arise in the oblate state (c) and triaxial state (d), respectively.

\subsection{Energy levels}

In this section, 
we describe the results of GCM calculations performed by superposing the obtained wave functions
on the $\beta$-$\gamma$ plane for negative- and positive-parity states.

\begin{figure}[t]
\centering
	\includegraphics[angle=-90, width=8.6cm, bb=310 54 546 402, clip]{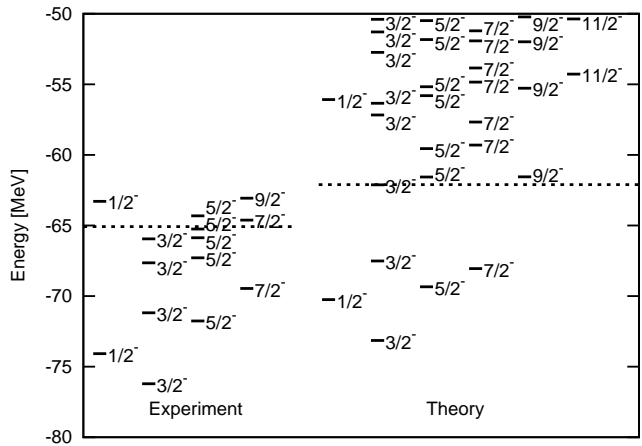}
	\caption{Energy levels of the negative-parity states in $^{11}$B.
	Four columns on the left are the experimental data and six columns on the right are the calculated results.
	The dotted lines in the left and right show 
	the experimental and theoretical $2\alpha + t$ threshold energies, respectively.}
\label{Energy_level_11B_-}
\end{figure}

\begin{figure}[t]
\centering
	\includegraphics[angle=-90, width=8.6cm, bb=300 54 556 402, clip]{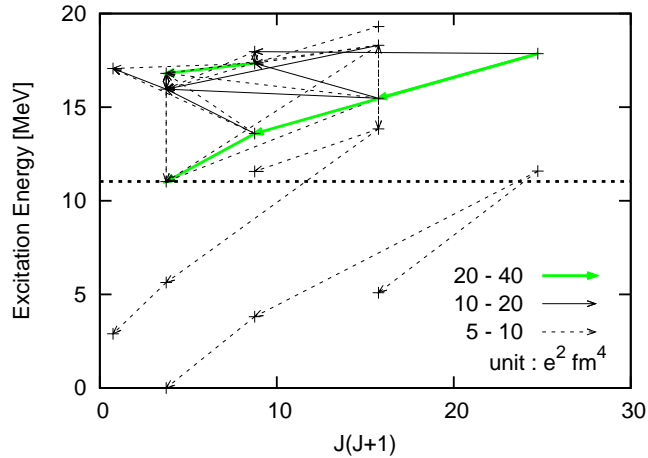}
	\caption{The calculated negative-parity states in $^{11}$B against the angular momentum $J(J+1)$ with $E2$ transition strengths.
		$5.0$ $e^{2}$fm$^{4} < B(E2) \le 10.0$ $e^{2}$fm$^{4}$,
		$10.0$ $e^{2}$fm$^{4} < B(E2) \le 20.0$ $e^{2}$fm$^{4}$, and
		$20.0$ $e^{2}$fm$^{4} < B(E2) \le 40.0$ $e^{2}$fm$^{4}$ transitions are described by 
		broken, black solid, and green bold solid arrows, respectively.
		The dotted line shows theoretical $2\alpha + t$ threshold energy.}
\label{J_vs_Energy_11B_-}
\end{figure}

First, we describe the results for the negative-parity states.
We show the calculated negative-parity energy levels in Fig.~\ref{Energy_level_11B_-}
as well as the experimental levels.
In the four columns on the left, we display the experimental energy levels 
for all the negative-parity assigned states \cite{Ajzenberg_A=11-12_90,Yamaguchi_11B_11}.
In the six columns on the right, the theoretical levels are illustrated.
In Fig.~\ref{J_vs_Energy_11B_-}, we plot the negative-parity energy levels 
as functions of the angular momentum $J(J+1)$ with $E2$ transition strengths.
We also show the calculated $E2$ transition strengths, isoscalar monopole transition strengths, and root-mean-square radii 
with experimental data \cite{Ajzenberg_A=11-12_90,Ozawa_rmsRadius_01,Kawabata_11B_07} 
in Tables~\ref{E2_11B_-}, \ref{E0_11B_-}, and \ref{Radius_11B_-}, respectively.
Our calculated results agree with experimental ones reasonably.

\begin{table}[t]
	\caption{Electromagnetic transition strengths $B(E2)$ for the negative-parity states in $^{11}$B.
	The unit is $e^{2}$fm$^{4}$.}
	\label{E2_11B_-}
	\centering
	\begin{tabular}[t]{ccc} \hline \hline
\multirow{2}{*}{Transition} & \multicolumn{2}{c}{Strength} \\ \cline{2-3}
 & Theory & Experiment \\ \hline
$5/2^{-}_{1} \rightarrow 3/2^{-}_{1}$ &  9.2 & $14 \pm 3$ \\
$7/2^{-}_{1} \rightarrow 3/2^{-}_{1}$ &  1.3 & $1.9 \pm 0.4$ \\
$5/2^{-}_{2} \rightarrow 3/2^{-}_{1}$ &  0.4 & $1.0 \pm 0.7$ \\
$3/2^{-}_{2} \rightarrow 1/2^{-}_{1}$ &  6.7 & $4 \pm 3$ \\
	\hline \hline
	\end{tabular}
\end{table}

\begin{table}[t]
	\caption{Isoscalar monopole transition strengths $B(E0;IS)$ for the negative-parity states in $^{11}$B.
	The unit is fm$^{4}$.}
	\label{E0_11B_-}
	\centering
	\begin{tabular}[t]{ccc} \hline \hline
\multirow{2}{*}{Transition} & \multicolumn{2}{c}{Strength} \\ \cline{2-3}
 & Theory & Experiment \\ \hline
$3/2^{-}_{1} \rightarrow 3/2^{-}_{2}$ &  2.5 & $< 9$ \\
$3/2^{-}_{1} \rightarrow 3/2^{-}_{3}$ &  150 & $96 \pm 16$ \\
	\hline \hline
	\end{tabular}
\end{table}

\begin{table}[t]
	\caption{Root-mean-square radii for mass distributions of the negative-parity states in $^{11}$B. The unit is fm.}
	\label{Radius_11B_-}
	\centering
	\begin{tabular}{ccc} \hline \hline
\multirow{2}{*}{State} & \multicolumn{2}{c}{Radius} \\ \cline{2-3}
 & Theory & Experiment \\ \hline
	$3/2^{-}_{1}$ & 2.29 & $2.09 \pm 0.12$ \\
	$3/2^{-}_{2}$ & 2.46 & \\
	$3/2^{-}_{3}$ & 2.65 & \\ 
	\hline \hline
	\end{tabular}
\end{table}

With help of $E2$ strengths and analysis of overlaps with basis wave functions, 
we here describe features of ground and excited states and band structures in the GCM results.
The calculated low-lying states have large overlaps with the basis wave functions in the small deformation region.
For instance, the $3/2^{-}_{1}$ state has 87\% overlap with the energy minimum state at 
$(\beta \cos \gamma, \beta \sin \gamma) = (0.33, 0.13)$ (Fig.~\ref{density_11B_-}(a)) in the $3/2^{-}$ energy surface.
As mentioned before, this state has the intermediate feature
between the shell-model structure and the cluster structure.
For the low-lying states, the calculated $E2$ strengths are reasonable compared with 
the experimental values though the level ordering is somehow in disagreement with the experimental one.

In the high-lying states above $-65$ MeV,
we obtain various developed cluster states having significant overlaps with the basis wave functions in the large $\beta$ regions,
such as Fig.~\ref{density_11B_-}(b), (c), and (d).
In particular, the $3/2^{-}_{3}$ state, which is considered to have a dilute cluster structure with a $2\alpha + t$ configuration,
is described by the linear combination of various $2\alpha+t$ spatial configurations.
In the next section, we discuss the relation between the $3/2^{-}_{3}$ state in $^{11}$B and the $0^{+}_{2}$ state in $^{12}$C
comparing these GCM amplitudes on the $\beta$-$\gamma$ plane.

For the $3/2^{-}_{3}$, $5/2^{-}_{3}$, $7/2^{-}_{3}$, and $9/2^{-}_{3}$ states, 
the $E2$ transition strengths are significantly large as 20-30 $e^{2}$fm$^{4}$ 
(See Fig.~\ref{J_vs_Energy_11B_-}), and therefore, we consider these states as members of a band 
starting from the band head $3/2^{-}_{3}$ state.
In the next section, we discuss the correspondence between this band and 
the experimental band suggested in Ref.~\cite{Yamaguchi_11B_11}.

\begin{figure}[t]
\centering
	\includegraphics[angle=-90,width=8.6cm, bb=300 54 546 402, clip]{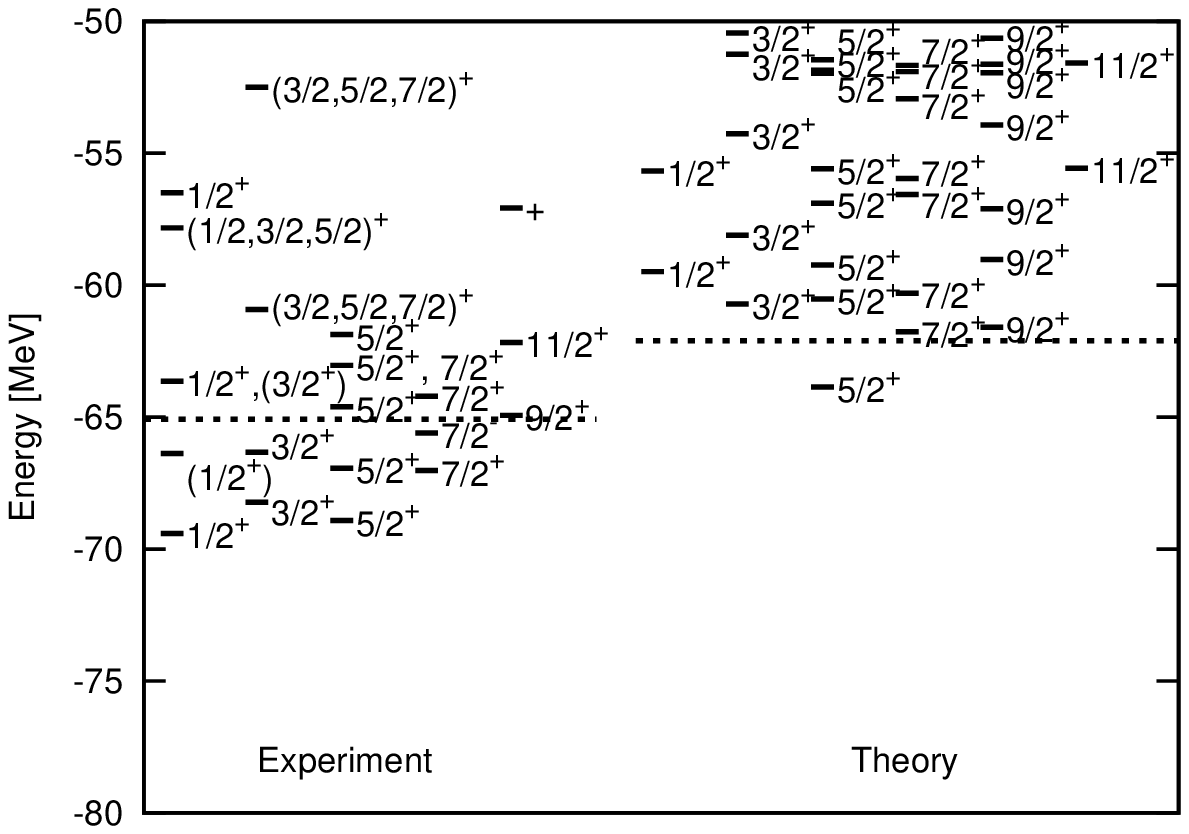}
	\caption{Energy levels of the positive-parity states in $^{11}$B.
	Five columns on the left are the experimental data and six columns on the right are the calculated results.
	The dotted lines in the left and right show 
	the experimental and theoretical $2\alpha + t$ threshold energies, respectively.}
\label{Energy_level_11B_+}
\end{figure}

Next, we describe the results for the positive-parity states.
We show the calculated positive-parity energy levels in Fig.~\ref{Energy_level_11B_+}
as well as the experimental levels.
In the five columns on the left, we display the experimental energy levels 
for all the positive-parity assigned states \cite{Ajzenberg_A=11-12_90}.
The theoretical levels are illustrated in the six columns on the right.

For the calculated $3/2^{+}_{1}$, $5/2^{+}_{1}$, $5/2^{+}_{2}$, $7/2^{+}_{1}$, $7/2^{+}_{2}$, $9/2^{+}_{1}$, $9/2^{+}_{2}$,
and $11/2^{+}_{1}$ states,
we found that these states are constructed dominantly from the AMD wave function
at $(\beta \cos \gamma,\beta \sin \gamma)=(0.60,0.09)$ (Fig.~\ref{density_11B_+}(a)) 
which is the energy minimum of the $5/2^{+}$ energy surface.

The calculated $1/2^{+}_{1}$, $3/2^{+}_{2}$, $5/2^{+}_{3}$, $7/2^{+}_{4}$, $9/2^{+}_{2}$, and $11/2^{+}_{2}$ states 
have almost 50\% overlap with the AMD base 
at $(\beta \cos \gamma,\beta \sin \gamma) = (1.10,0.00)$ (Fig.~\ref{density_11B_+}(b)).
As will be discussed later, this basis wave function can be considered to have 
the molecular 2$\alpha$+$p(\sigma_{1/2})$+2$n((\sigma_{1/2})^{2})$ structure
in analogy to molecular orbital structures in $^{10}$Be.
The level spacings of these states show the $K^{\pi}=1/2^{+}$ rotational pattern.
Moreover, the calculated $E2$ transition strengths between these states listed in Tables~\ref{E2_11B_+} 
show a feature of the $K^{\pi}=1/2^{+}$ rotational band.
Namely, the transitions in the groups ($1/2^{+}_{1}$, $5/2^{+}_{3}$, and $9/2^{+}_{3}$) and 
($3/2^{+}_{2}$, $7/2^{+}_{4}$, and $11/2^{+}_{2}$) are rather strong, 
while those between the groups are weak.
Therefore, we regard the $1/2^{+}_{1}$, $3/2^{+}_{2}$, $5/2^{+}_{3}$, $7/2^{+}_{4}$, $9/2^{+}_{3}$, 
and $11/2^{+}_{2}$ states as the band members of the $K^{\pi}=1/2^{+}$ rotational band.

Other states have no specific structure and are difficult to be classified as band members.

In the present calculation, the $1/2^{+}_{1}$ state, which is experimentally known to be the lowest positive-parity state, is missing.
It is expected to have $1p$-$1h$ configuration with one proton in the spatial extending $1s_{1/2}$ orbital.
Unfortunately, the present framework may not be suitable to describe the spatial extent of $1s_{1/2}$ orbital because 
the width parameter is taken to be a common value for all the single-particle Gaussian wave functions.

\begin{table}[t]
	\caption{$B(E2)$ for the positive-parity linear states in $^{11}$B.
	The unit is $e^{2}$fm$^{4}$.}
	\label{E2_11B_+}
	\centering
	\begin{tabular}[t]{cc} \hline \hline
Transition & Strength \\ \hline
$11/2^{+}_{2} \rightarrow 7/2^{+}_{4}$ &  154.8 \\
$7/2^{+}_{4} \rightarrow 3/2^{+}_{2}$ &  85.6 \\
& \\
$9/2^{+}_{3} \rightarrow 5/2^{+}_{3}$ &  84.1 \\
$5/2^{+}_{3} \rightarrow 1/2^{+}_{1}$ &  48.6 \\
& \\
$11/2^{+}_{2} \rightarrow 9/2^{+}_{3}$ &  1.4 \\
$9/2^{+}_{3} \rightarrow 7/2^{+}_{4}$ &  6.7 \\
$7/2^{+}_{4} \rightarrow 5/2^{+}_{3}$ &  3.8 \\
$5/2^{+}_{3} \rightarrow 3/2^{+}_{2}$ &  0.9 \\
$3/2^{+}_{2} \rightarrow 1/2^{+}_{1}$ &  10.7 \\
	\hline \hline
	\end{tabular}
\end{table}

\section{Discussion}\label{discussion}

\subsection{Negative-parity band from the $3/2^{-}_{3}$ state}

\begin{figure}[t]
\centering
	\includegraphics[angle=-90, width=8.6cm, bb=300 54 556 402, clip]{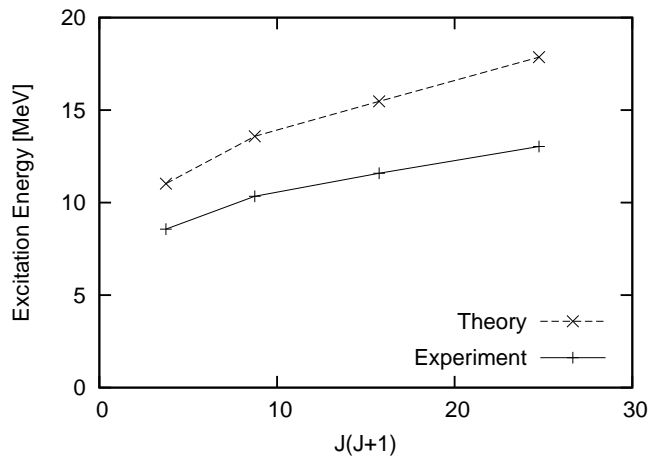}
	\caption{Comparison between the calculated band ($3/2^{-}_{3}$, $5/2^{-}_{3}$, $7/2^{-}_{3}$, and $9/2^{-}_{3}$) and
	the experimental band suggested in Ref.~\cite{Yamaguchi_11B_11}.}
	\label{comparison_exp_cal_11B}
\end{figure}

We here discuss the negative-parity band starting from the $3/2^{-}_{3}$ state.
As mentioned, the present result suggest the negative parity band consisting 
of the $3/2^{-}_{3}$, $5/2^{-}_{3}$, $7/2^{-}_{3}$, and $9/2^{-}_{3}$ states (Fig.~\ref{J_vs_Energy_11B_-}).
In the recent experiment of alpha resonant scattering on $^{7}$Li \cite{Yamaguchi_11B_11}, 
new cluster states were observed, and the negative-parity states 
at 8.56 MeV ($3/2^{-}$), 10.34 MeV ($5/2^{-}$), 11.59 MeV ($7/2^{-}$), and 13.03 MeV ($9/2^{-}$) are assigned to be band members.
We compare the experimental and calculated band in Fig.~\ref{comparison_exp_cal_11B}.
Although the calculated excitation energies are higher than experimental ones by 2.5 - 4.5 MeV, 
systematics of the level structure, in particular, the small level spacings correspond well to the experimental ones.
Moreover, the present result suggest developed cluster structures in this band
and these states probably have large $\alpha$ decay widths. 
It supports again that the calculated band can be assigned to the experimental band for which large $\alpha$ decay widths were suggested.
To make the correspondence of the calculated states to the experimental observed ones clearer,  
theoretical estimation of the partial decay widths of the excited states above the $2\alpha + t$ threshold 
is a remaining future problem.

\begin{figure}[t]
\centering
	\begin{tabular}{cc}
	\includegraphics[width=8.0cm, bb=7 11 592 301, clip]{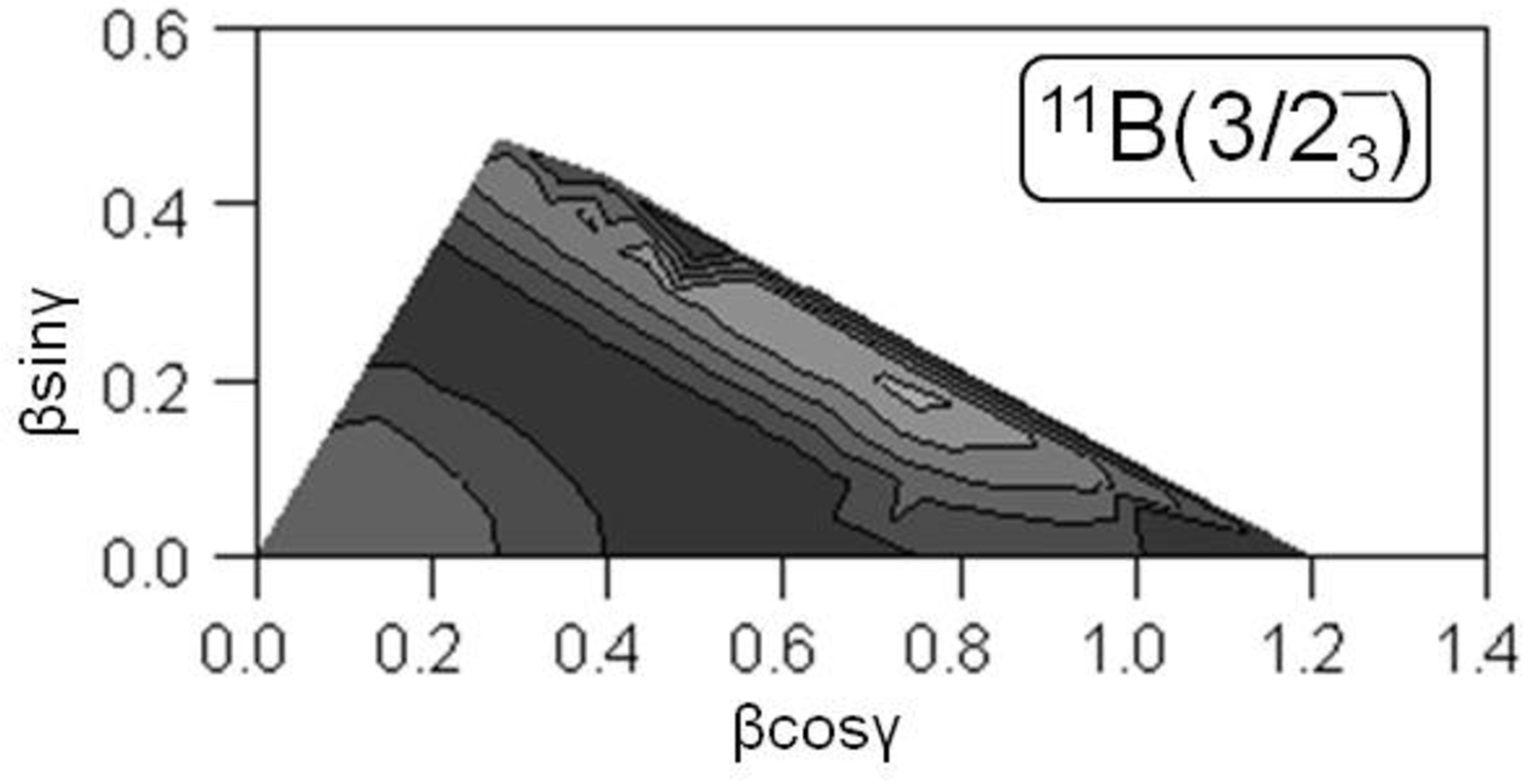} &
	\includegraphics[width=0.85cm, bb=27 2 66 165, clip]{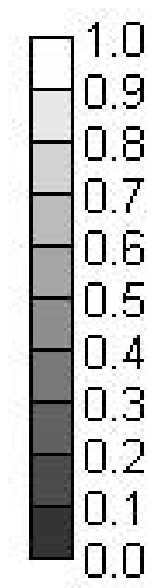} \\
	\includegraphics[width=8.0cm, bb=7 11 592 301, clip]{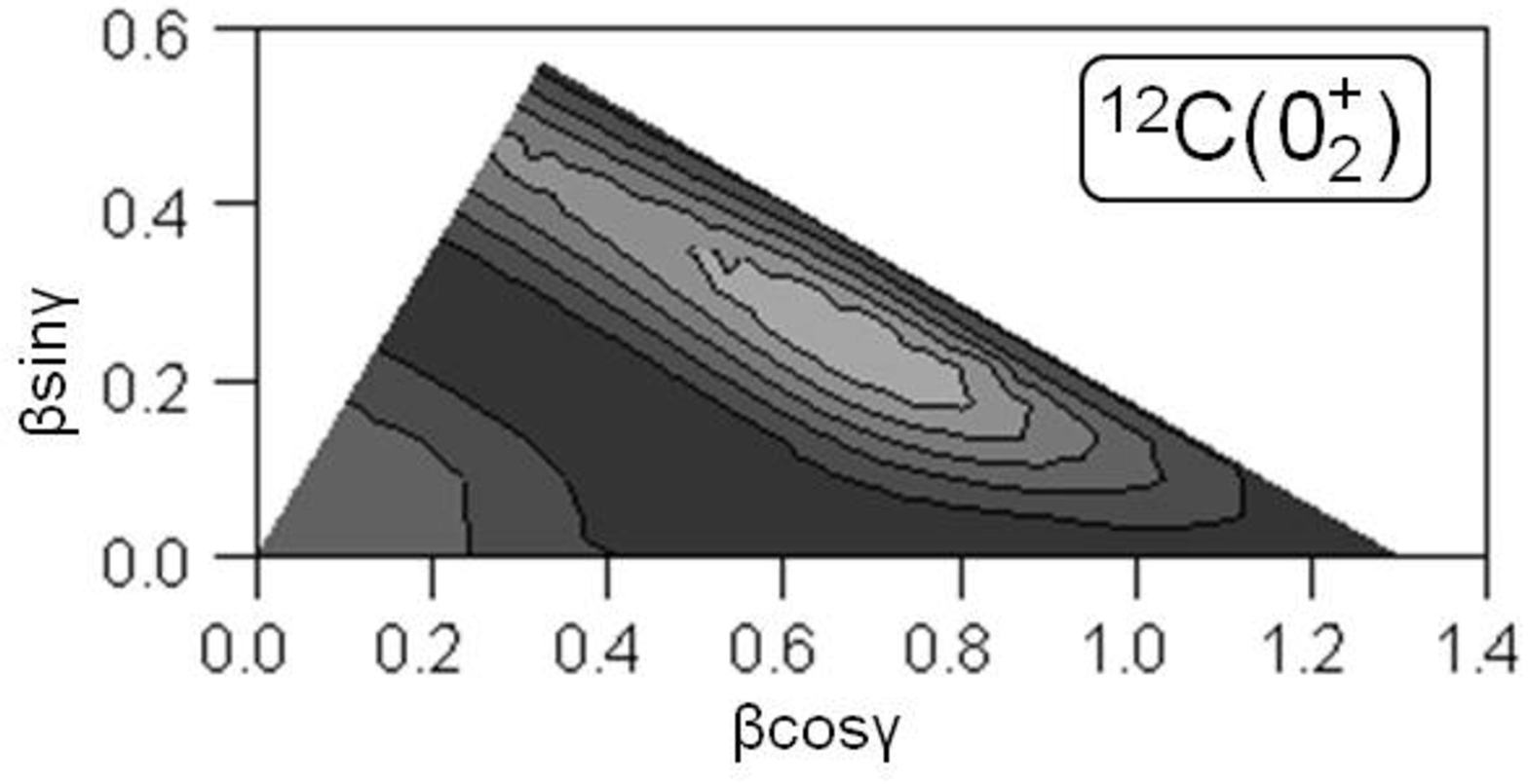}
	\end{tabular}
	\caption{GCM amplitudes for the $3/2^{-}_{3}$ states of $^{11}$B and the $0^{+}_{2}$ states of $^{12}$C.}
	\label{overlaps}
\end{figure}

Cluster structure of the band head state, the $3/2^{-}_{3}$ state, of this band 
has been attracting a special attention in association with the $0^{+}_{2}$ state in $^{12}$C.
In the earlier works \cite{En'yo_11B_07,Kawabata_11B_07}, 
it is suggested that the $3/2^{-}_{3}$ state has a similar feature to the $0^{+}_{2}$ state in $^{12}$C.
These states in $^{12}$C and $^{11}$B are considered to have dilute cluster structures 
with 3$\alpha$ and $2\alpha + t$ configurations, respectively.
To see the similarity of the cluster feature between these states, 
we compare the GCM amplitude for the $3/2^{-}_{3}$ states of $^{11}$B
with that for the $0^{+}_{2}$ state of $^{12}$C in Fig.~\ref{overlaps}. 
The results for the $0^{+}_{2}$ states in $^{12}$C are calculations with 
the $\beta$-$\gamma$ constraint AMD + GCM taken from Ref.~\cite{Suhara_AMD_10}.
Fragmentation of the GCM amplitudes for these states is very similar to each other.
Both GCM amplitudes spread over the broad $\gamma$ area of the large $\beta$ region.
In this area of the large $\beta$ region, the basis wave functions have various configurations of the developed clusters.
The feature of the GCM amplitudes indicates that the $3/2^{-}_{3}$ state in $^{11}$B is described  
by the linear combination of various $2\alpha+t$ configurations in the same manner as the $0^{+}_{2}$ state in $^{12}$C,
which is also described by the linear combination of various 3$\alpha$ configurations.
That is, when one $\alpha$ cluster is replaced to one $t$ cluster,
the structure of the $3/2^{-}_{3}$ state in $^{11}$B can be regarded as a very similar state to the $0^{+}_{2}$ state in $^{12}$C. 
Our calculation is consistent with earlier works.

Let us turn again to the level structure of the negative-parity band starting from the $3/2^{-}_{3}$ state.
As shown in Fig.~\ref{comparison_exp_cal_11B}, 
the energy positions of the $5/2^{-}_{3}$, $7/2^{-}_{3}$, and $9/2^{-}_{3}$ states
satisfy the $E_{\rm rot} \propto J(J+1)$ rule of the rigid rotor model, 
while that of the $3/2^{-}_{3}$ deviate from this rule and it is lower than the systematic line. 
This feature is seen in both the calculated and experimental levels.
We here propose a possible reason for the lowering $3/2^{-}_{3}$ 
while focusing on its dilute cluster structure as follows. 
In the $3/2^{-}_{3}$ state, two $\alpha$ and $t$ clusters are weakly interacting, 
and it may not have a rigid structure. 
Therefore, energy cost for rotation of this state can be relatively large. 
As the angular momentum increase, the structure may change from weakly interacting clusters 
to a somehow rigid structure with a specific shape resulting in a larger moment of inertia. 
By a consequence of the change of moment of inertia, the plot of the energy levels with respect to $J(J+1)$ shows a kink.
A similar feature was discussed for the band in $^{16}$O, 
which are considered to start from the $0^{+}_{6}$ state, 
a candidate of the $\alpha$ condensation having weakly interacting four $\alpha$ clusters \cite{Funaki_16O_11}.

\subsection{Molecular orbital structures in positive-parity states}

As mentioned before, 2$\alpha$ core structures are found in positive-parity states. 
By considering orbitals around the 2$\alpha$ core, 
we can describe structures of positive-parity states in the molecular orbital picture.
The molecular orbital picture was proposed to describe systems of a $2\alpha$ core 
with valence neutron(s) \cite{Okabe_9Be_77,Okabe_9Be_79,Seya_MO_81,OERTZEN,OERTZENa,Oertzen-rev}. 
Indeed, low-lying states of neutron-rich Be isotopes are successfully described 
by the molecular orbital pictures \cite{Seya_MO_81,OERTZEN,OERTZENa,Oertzen-rev,Dote:1997zz,ENYObe10,Itagaki_10Be_00,Itagaki_10Be_02, Arai01,Ito_10Be_04,Ito_10Be_06}. 

In a 2$\alpha$ system with valence neutrons,
molecular orbitals are formed by a linear combination of $p$ orbits around two $\alpha$ clusters, 
and valence neutrons occupy the molecular orbitals such as $\pi_{3/2}$ and $\sigma_{1/2}$ orbitals. 
The $\pi_{3/2}$ orbital ($J^{\pi}=3/2^{-}$) 
spreads in a direction perpendicular to the axis between $\alpha$ clusters,
while the $\sigma_{1/2}$ orbital ($J^{\pi}=1/2^{+}$) 
spreads parallel to the axis between $\alpha$ clusters.
In the ground band of $^{10}$Be, valence neutrons have the $(\pi_{3/2})^{2}$ configuration.
While, in the excited states of $^{10}$Be, 
developed cluster structures with other configurations of valence neutrons appear.
For instance, $K^{\pi}=1^{-}_{1}$ and $K^{\pi}=0^{+}_{2}$ bands are understood by 
$\pi_{3/2} \sigma_{1/2}$ and $(\sigma_{1/2})^{2}$ configurations, respectively. 
A molecular orbital model was also applied to B isotopes as well as Be isotopes by Seya {\it et al.} \cite{Seya_MO_81}.

We here consider a 2$\alpha$ core with surrounding a proton and two neutrons 
in the molecular orbitals for $^{11}$B and discuss structures of positive-parity states 
in terms of the molecular orbitals in association with molecular orbital structures in $^{10}$Be.

First, we consider the calculated $3/2^{+}_{1}$, $5/2^{+}_{1}$, $5/2^{+}_{2}$, $7/2^{+}_{1}$, 
$7/2^{+}_{2}$, $9/2^{+}_{1}$, $9/2^{+}_{2}$, and $11/2^{+}_{1}$ states.
The main component of those states is the wave function at 
$(\beta \cos \gamma,\beta \sin \gamma)=(0.60,0.09)$ (Fig.~\ref{density_11B_+}(a)),
which has a 2$\alpha$+$p$+2$n$ configuration.
The motion of one valence proton of this wave function can be regarded as the $\pi_{3/2}$ molecular orbitals 
because one valence proton attach the side of two $\alpha$ clusters and 
two valence neutrons can be interpreted as occupying $\pi_{3/2}$ and $\sigma_{1/2}$ molecular orbitals.
The density distribution of neutrons has a banana shape, 
which is constructed from the combination of the $\pi_{3/2}$ and $\sigma_{1/2}$ molecular orbitals.
As the results, $3/2^{+}_{1}$, $5/2^{+}_{1}$, $5/2^{+}_{2}$, $7/2^{+}_{1}$, $7/2^{+}_{2}$, $9/2^{+}_{1}$, $9/2^{+}_{2}$,
and $11/2^{+}_{1}$ states can be interpreted as 2$\alpha$+$p(\pi_{3/2})$+2$n(\pi_{3/2} \sigma_{1/2})$.
This neutron configuration is similar to that of the $K^{\pi}=1^{-}_{1}$ band in $^{10}$Be.

Next, we discuss the $K^{\pi}=1/2^{+}$ band in $^{11}$B.
The $K^{\pi}=1/2^{+}$ band is constructed dominantly from the wave function 
at $(\beta \cos \gamma, \beta \sin \gamma)=(1.10, 0.00)$ (Fig.~\ref{density_11B_+}(b)).
In this wave function, protons and neutrons have the elongate structure parallel to the axis between two $\alpha$ clusters.
This structure has, therefore, the strong coupling feature and 
it is proper to be interpreted in terms of the molecular orbital structure.
This elongate structure is consistent with the $\sigma_{1/2}$ orbital.
Therefore, this wave function can be regarded as 2$\alpha$+$p(\sigma_{1/2})$+2$n((\sigma_{1/2})^{2})$, 
where a proton and two neutrons occupy the $\sigma_{1/2}$ orbital.
The motion of valence neutrons in $K^{\pi}=1/2^{+}$ is similar to 
that of valence neutrons in the $0^{+}_{2}$ state of $^{10}$Be.

As the results, we find a good correspondence of the intrinsic structure 
of positive-parity states in $^{11}$B to the excited states in $^{10}$Be.
In the excited states of $^{10}$Be, molecular 2$\alpha$+2$n$ structures with
$\pi_{3/2}\sigma_{1/2}$ and $(\sigma_{1/2})^{2}$ configurations construct rotational bands.
Also in the positive-parity states of $^{11}$B, there are two molecular orbital configurations 
such as $p(\pi_{3/2})$+2$n(\pi_{3/2} \sigma_{1/2})$ and $p(\sigma_{1/2})$+2$n((\sigma_{1/2})^{2})$.
That is, the excited states of both nuclei can be described by molecular orbital structures with two $\alpha$ clusters.
Note that valence neutrons occupy the same molecular orbitals
such as $\pi_{3/2}\sigma_{1/2}$ and $(\sigma_{1/2})^{2}$ in both nuclei.
It suggests that the molecular orbital structures of $^{11}$B can be composed by 
an additional proton and $^{10}$Be with the corresponding molecular orbital structures.

Here, we discuss whether or not other configurations of molecular orbitals exist in $^{11}$B.
In the simple expectation from the molecular orbital model, other configurations for valence nucleons
can appear near or under the $2\alpha$+$p(\sigma_{1/2})$+2$n(\sigma_{1/2})^2$ states.
For example, the $2\alpha$+$p(\pi_{3/2})$+2$n(\sigma_{1/2})^2$ structure can appear in the negative-parity states
because the $\pi_{3/2}$ orbital has lower energy than the $\sigma_{1/2}$ orbital in a simple point of view.
However, in the $\beta$-$\gamma$ constraint AMD + GCM calculation, such the state does not appear.
We consider the reason for the absence is less correlation energy among nucleons in molecular orbitals.
In the $2\alpha$+$p(\pi_{3/2})$+2$n(\sigma_{1/2})^2$ states,
since an valence proton and valence neutrons occupy the different orbitals they have small spatial overlaps; 
therefore, valence nucleons gain less correlation energy.
However, in the molecular orbital states which appear in the $\beta$-$\gamma$ constraint AMD + GCM calculation,
three valence nucleons (a proton and two neutrons) occupy the same orbital and they gain much correlation energy.
In other words, the molecular orbital states in which valence proton and neutrons do not occupy same orbitals are unfavored and 
cannot appear in the low energy region.

\subsection{Coexistence of shell-model, $2\alpha + t$, and molecular orbital structures}

Let us here discuss the coexistence of various structures in $^{11}$B.
What we find in the present work is two-types of cluster structures, 
three-body $2\alpha + t$ cluster and molecular orbital structures as well as shell-model structures. 
$2\alpha + t$ cluster structures in $^{11}$B correspond well to 3$\alpha$ cluster structures in $^{12}$C, 
while molecular orbital structures can be associated with those in $^{10}$Be. 
It should be emphasized that the coexistence of three-body $2\alpha + t$ cluster and molecular orbital structures is 
one of the unique features of $^{11}$B.

In $^{11}$B system, shell-model structures are seen in the ground and low-lying states, 
while in the highly excited states near or above the $2\alpha + t$ threshold, 
three-body $2\alpha + t$ structures and also molecular orbital structures with a 2$\alpha$ core are found. 
In the molecular orbital structures, the 2$\alpha$ core is formed and 
three valence nucleons (a proton and two neutrons) are moving in the mean field, 
i.e., the molecular orbitals around the $2\alpha$. 
It might seem to be contrast to the $2\alpha + t$ structures 
where a $t$ cluster is formed by correlating three nucleons. 
The reason why these two kinds of cluster structures coexist in a similar energy region 
can be understood by fragility of a $t$ cluster. 
The binding energy of a $t$ cluster is only 8.5 MeV and much smaller than that of an $\alpha$ cluster. 
It means that a $t$ cluster can easily break up into three nucleons. 
In molecular orbital structures, three valence nucleons are in the mean field 
to gain potential energy from the $2\alpha$ core. 
The potential energy gain from the core can compensate the energy loss of the binding energy of a $t$ cluster. 
As a result, $2\alpha + t$ cluster and molecular orbital structures coexist in $^{11}$B.

As already mentioned before, the coexistence of shell-model structures and three-body cluster structures 
occurs in $^{11}$B in a similar way to the coexistence of cluster and shell-model structures in $^{12}$C. 
Then, we can conclude that shell-model structures, $2\alpha + t$ cluster structures, 
and molecular orbital ones coexist in the ground and excited states of $^{11}$B. 
This is the new coexistence phenomenon peculiar to $^{11}$B.

\section{Summary}\label{summary}

We investigated structures of excited states in $^{11}$B with the method of $\beta$-$\gamma$ constraint AMD + GCM.
We showed the calculated results for the energy levels, 
$E2$ transition strengths, isoscalar monopole transition strengths, and a root-mean-square radius, 
which are in reasonable agreements with experimental data.
The present results suggest that the ground and low-lying states have the shell-model structures.
In the excited states, well-developed cluster structures are found in the negative- and positive-parity states.

By analyzing the $E2$ transition strengths as well as the GCM amplitudes, 
we assigned the negative-parity band starting from $3/2^{-}_{3}$, in which
a $2\alpha + t$ structure develops well.
This band is the candidate for a band which was suggested in the recent experiment of alpha resonant scattering on $^{7}$Li.
Systematics of the level structure in this band, in particular, the small level spacings correspond well to the experimental ones.
In the experiment, this band is constructed by the states with large $\alpha$ decay widths.
To make the correspondence of the calculated states to the experimental observed ones clearer,  
theoretical estimation of the partial decay widths of the excited states above the $2\alpha + t$ threshold 
is a remaining future problem.

In the negative-parity states, the present results suggest a good correspondence of the intrinsic structure 
with the positive-parity states of $^{12}$C as suggested in previous studies.
For instance, we found the well-developed $2\alpha$+$t$ cluster structure of the $3/2^{-}_{3}$ state in $^{11}$B, 
which shows similar features to $3\alpha$ cluster structure of the $0^{+}_{2}$ state in $^{12}$C.
In the analysis of GCM amplitudes, we found that the intrinsic structure of the $3/2^{-}_{3}$ state in $^{11}$B 
does not have a rigid shape but it is expressed by a linear combination of basis wave functions
having various configurations of the developed clusters. 
The feature is quite similar to the $0^{+}_{2}$ state in $^{12}$C.

For the positive-parity states, the intrinsic structures are described in terms of the molecular orbital structure, 
and a correspondence of the positive-parity states in $^{11}$B to the excited states in $^{10}$Be were discussed.
The low-lying states in the positive parity states of $^{11}$B have
the 2$\alpha$+$p(\pi_{3/2})$+2$n(\pi_{3/2} \sigma_{1/2})$ structure, 
which is similar to the $K^{\pi}=1^{-}_{1}$ band in $^{10}$Be
(2$\alpha$+2$n(\pi_{3/2} \sigma_{1/2})$) except for the valence proton.
The $K^{\pi}=1/2^{+}$ band has the 2$\alpha$+$p(\sigma_{1/2})$+2$n((\sigma_{1/2})^{2})$ structure. 
This is also similar to the $K^{\pi}=0^{+}_{2}$ band in $^{10}$Be
(2$\alpha$+2$n(\sigma_{1/2})^{2}$).

What we found in the present study of $^{11}$B is the coexistence of shell-model, $2\alpha$+$t$, and molecular orbital structures.
Shell-model structures are seen in the ground and low-lying states, while
in the highly excited states near or above the $2\alpha + t$ threshold, 
three-body $2\alpha + t$ structures and also molecular orbital structures are found.
This is a new coexistence phenomenon peculiar to $^{11}$B.
It is a future problem to investigate other nuclei from the point of view whether or not this coexistence appears.

\section*{Acknowledgments}

We truly appreciate the useful discussions of Prof. T. Kawabata and Dr. H. Yamaguchi.
The computational calculations of this work were performed by supercomputers in YITP and KEK. 
This work was supported by the YIPQS program in YITP.
It was also supported by the Grant-in-Aid for the Global COE Program 
``The Next Generation of Physics, Spun from Universality and Emergence" 
from the Ministry of Education, Culture, Sports, Science and Technology (MEXT) of Japan,
and Grant-in-Aid for Scientific Research from JSPS.

\end{document}